\documentclass[twoside]{article}
\usepackage{amsmath,amsfonts,amssymb,mathrsfs}
\usepackage{qic,epsfig}
\usepackage{tikz}
\usepackage{cite}
\usepackage{color}
\usepackage{booktabs}
\usepackage{cases}

\renewcommand{\theequation}{\arabic{section}.\arabic{equation}}

\newtheorem{prop}{Proposition}
\newtheorem{thm}{Theorem}
\newtheorem{cor}{Corollary}
\newtheorem{lem}{Lemma}
\newtheorem{defn}{Definition}

\renewenvironment{proof}{\par\noindent{\bf Proof.}}{$\quad\Box$\par}
\newcommand{\ket}[1]{| #1 \rangle}
\newcommand{\bra}[1]{\langle #1 |}
\newcommand{\braket}[2]{\langle #1 | #2 \rangle}

\begin{document}
 \date{}

\setlength{\textheight}{8.0truein}    

\runninghead{Fault Tolerance Quantum Computation in Quotient
Algebra Partition} { }

\normalsize\textlineskip \thispagestyle{empty}
\setcounter{page}{1}

\vspace*{0.88truein}

\alphfootnote

\fpage{1}

\centerline{\bf Every Action in Every Code is Fault Tolerant: }
\centerline{\bf Fault Tolerance Quantum Computation in  Quotient
Algebra Partition } \vspace*{0.035truein}
\centerline{\footnotesize Zheng-Yao Su\footnote{Email:
zsu@nchc.narl.org.tw}\hspace{.15cm}and Ming-Chung Tsai}
\centerline{\footnotesize\it National Center for High-Performance
 Computing,}
 \centerline{\footnotesize\it National Applied Research Laboratories,
 Taiwan, R.O.C.}

\vspace*{0.21truein}

\abstracts{In the framework quotient algebra partition,
 a general methodology is introduced to construct
 fault tolerant encodes for an arbitrary action in an error-correcting code.}{}{}

\vspace*{1pt}\textlineskip

 \section{Introduction}\label{secintro}
 Quantum computer is proven theoretically to possess
 greater computational power than its classical counterpart~\cite{ShorAlg,GroverAlg}.
 This prowess is expected to be presented in a reliable large-scale
 hardware against an error-prone environment.
 Fault tolerance computation is essential towards this aim,
 during which the scheme of error correction plays the central role~\cite{PeterShor,KnillLaflamme,GottesmanStab}.
 In current attempts of realizing fault tolerance quantum computation,
 transversal encodes are the main focus~\cite{TerhalNature,LfalammeFTQC}.

 In this work,
 fault tolerant encodes for an action are constructed in the framework
 called {\em Quotient Algebra Partition (QAP)}.
 Originally sketched in 2005~\cite{OriginQAPSu},
 a QAP is a partition over a Lie algebra
 that consists of abelian subspaces closed under
 commutation relations.
 Inheriting algebraic features and geometric properties of Lie algebras and Lie groups~\cite{Helgason,Knapp},
 the QAP structure manifests their {\em combinatorial traits}~\cite{OriginQAPSu,QAPSu0,QAPSu1,QAPSuTsai1,QAPSuTsai2}.
 Through an isomorphism,
 the code $[n,k,\hspace{1pt}{\cal C}]$ of a stabilizer ${\cal C}$
 acquires a partition from the framework
 and is viewed a {\em quantum extension} of a Hamming code.
 Given the code $[n,k,\hspace{1pt}{\cal C}]$,
 the error correction is depicted in the partition structure.
 Resorting to a spinor-to-spinor transformation
 $Q^{\dag}_{en}\in{SU(2^n)}$,
 the partition $[n,k,\hspace{1pt}{\cal C}]$
 is mapped to the so-called {\em intrinsic coordinate}.
 In this special coordinate,
 multiple choices of the fault tolerant version
 $\hat{U}\in{SU(2^n)}$
 are composed for a $k$-qubit action.
 Then, via $Q_{en}$, {\em i.e.}, an encoding,
 a fault tolerant encode
 $U=Q_{en}\hspace{1pt}\hat{U}\hspace{1pt}Q^{\dag}_{en}$
 is delivered to the code.
 This article is self-contained,
 and explicit examples are offered pedagogically in~\cite{SuTsaiOptFTQC}.

 \section{QAP Structure}\label{secQAPtoStabC}
  To portray the structure QAP,
  it needs a language of convenience for spinors~\cite{QAPSu0,QAPSu1}.
  \vspace{6pt}
  \begin{defn}\label{s-rep}
  With an encapsulation of Pauli matrices
  \begin{align}\label{eqsrep1qubit}
 {\cal S}^{\epsilon_i}_{a_i}=(\ket{0}\bra{a_i}+(-1)^{\epsilon_i}\ket{1}\bra{1+a_i}),
 \end{align}
 the $s$-representation of an $n$-qubit spinor is written in the form
 \begin{align}\label{eqSrepinapp}
 {\cal S}^{\zeta}_{\alpha}={\cal S}^{\epsilon_{1}\epsilon_2\ldots
 \epsilon_{n}}_{a_{1} a_2\ldots a_{n}}
 ={\cal S}^{\epsilon_{1}}_{a_{1}}\otimes{\cal S}^{\epsilon_{2}}_{a_{2}}\ldots\otimes{\cal S}^{\epsilon_{n}}_{a_{n}},
 \end{align}
 where $\alpha$ is termed as the binary partitioning string or the bit string
 and $\zeta$ the phase string of ${\cal S}^{\zeta}_{\alpha}$,
 $a_i,\epsilon_i\in{Z_2}$ and $i=1,2,\cdots,n$.
  \end{defn}
  \vspace{6pt}
 Specifically, the Pauli matrices are rewritten as
 ${\cal S}^{0}_{\hspace{.81pt}0}=I$, ${\cal S}^{1}_{\hspace{.9pt}0}=\sigma_{3}$,
 ${\cal S}^{0}_{\hspace{.92pt}1}=\sigma_{1}$ and $-i{\cal S}^{1}_{\hspace{.92pt}1}=\sigma_{2}$.
 Two instances are given for example
 ${\cal S}^{010}_{101}=\sigma_{1}\otimes\sigma_{3}\otimes\sigma_{1}$ and
 $-i{\cal S}^{0100}_{0111}=I\otimes\sigma_{2}\otimes\sigma_{1}\otimes\sigma_{1}$.
 \vspace{6pt}
 \begin{lem}\label{lemproductof2spins}
 The multiplication of spinors
  ${\cal S}^{\hspace{.5pt}\zeta}_{\alpha}$ and ${\cal S}^{\eta}_{\beta}\in{su(2^n)}$
  turns into additions of binary strings
 \begin{align}\label{eq-multiple2Spinors}
  {\cal S}^{\hspace{.5pt}\zeta}_{\alpha}\cdot{\cal S}^{\eta}_{\beta}
  =(-1)^{\eta\cdot\alpha}{\cal S}^{\zeta+\eta}_{\alpha+\beta}.
 \end{align}
 \end{lem}
 \vspace{2pt}
 \begin{proof}
  The derivation is straightforward~\cite{QAPSu0}.
 In Eq.~\ref{eq-multiple2Spinors}, the addition of two binary strings is a {\em bitwise addition},
 and the exponent
 $\eta\cdot\alpha=\sum^{n}_{i=1}\sigma_{i} a_{i}$ is the {\em inner product} of two strings of the same length,
  $\eta=\sigma_1\sigma_2\cdots\sigma_{n}$ and $\alpha=a_1a_2\cdots a_n\in{Z^n_2}$.
 An immediate implication is the commutability of two generators
 \begin{align}\label{eqcommanticommspinors}
  {\cal S}^{\zeta}_{\alpha}{\cal S}^{\eta}_{\beta}
  =(-1)^{\eta\cdot\alpha+\zeta\cdot\beta}{\cal S}^{\eta}_{\beta}{\cal S}^{\zeta}_{\alpha}
  =
  \begin{cases}
      \hspace{9pt}{\cal S}^{\eta}_{\beta}{\cal S}^{\zeta}_{\alpha}    & \quad \text{if } \eta\cdot\alpha+\zeta\cdot\beta=0,2;\\
     -{\cal S}^{\eta}_{\beta}{\cal S}^{\zeta}_{\alpha}                & \quad \text{if } \eta\cdot\alpha+\zeta\cdot\beta=1,3.
  \end{cases}
 \end{align}
 The {\em parity condition} indicates that the two spinors commute if $\eta\cdot\alpha+\zeta\cdot\beta=0,2$
  or anticommute if $\eta\cdot\alpha+\zeta\cdot\beta=1,3$.
 \end{proof}
 \vspace{6pt}
 A spinor can act on states.
 \vspace{6pt}
 \begin{lem}\label{lemspinactsonSt}
 A spinor ${\cal S}^{\zeta}_{\alpha}$ maps a basis
 state of $n$ qubits into another
 \begin{align}\label{spinoronstate}
  {\cal S}^{\zeta}_{\alpha}\ket{\beta}
 =(-1)^{\zeta\cdot(\beta+\alpha)}\ket{\beta+\alpha}.
 \end{align}
 \end{lem}
 \vspace{2pt}
 \begin{proof}
 Similarly, refer to~\cite{QAPSu0} for its plain derivation.
 Being a generator of the algebra $su(2^n)$,
 the spinor ${\cal S}^{\zeta}_{\alpha}$
 is also a group action
 in $SU(2^n)$, thanks to the exponential mapping
 $i(-i)^{\zeta\cdot\alpha}{\cal S}^{\zeta}_{\alpha}=e^{i\frac{\pi}{2}(-i)^{\zeta\cdot\alpha}{\cal
 S}^{\zeta}_{\alpha}}$,
 {\em cf.} Lemma~\ref{lemS-rot}.
 \end{proof}
 \vspace{6pt}
 {\bf {\em Nota Bene}}\hspace{5pt}
  A spinor $(-i)^{\zeta\cdot\alpha}{\cal S}^{\zeta}_{\alpha}$,
  multiplied by the phase $(-i)^{\zeta\cdot\alpha}$, is {\em hermitian},
  {\em i.e.},\\
  $((-i)^{\zeta\cdot\alpha}{\cal S}^{\zeta}_{\alpha})^{\dag}=(-i)^{\zeta\cdot\alpha}{\cal S}^{\zeta}_{\alpha}$,
  and {\em involutory}
  $((-i)^{\zeta\cdot\alpha}{\cal S}^{\zeta}_{\alpha})^2={\cal S}^{\mathbf{0}}_{\mathbf{0}}$~\cite{QAPSu0,QAPSu1}.
 \vspace{6pt}
 \\
 To retain the hermiticity,
 the inner product $\zeta\cdot\alpha$ counts the number of
 occurences of the 1-qubit component ${\cal S}^1_1$ in an
 $n$-qubit spinor
 ${\cal S}^{\zeta}_{\alpha}=\bigotimes^n_{i=1}{\cal S}^{\epsilon_i}_{a_i}$,
 and the accumulated exponent $\zeta\cdot\alpha$ of the phase
 $(\pm i)^{\zeta\cdot\alpha}$ is modulo $4$.
 While, this phase is often
 omitted if no confusion arises.

 The equivalence of two string additions and a spinor product
 suggests an operation~\cite{QAPSu0,QAPSu1}. 
 \vspace{-6pt}
 \begin{defn}\label{defnbiaddtion}
  The bi-addition $\hspace{2pt}\diamond$ is an additive operation
  on the Lie algebra $su(2^n)$
 \begin{align}\label{eqBiadd}
 {\cal S}^{\zeta}_{\alpha} \diamond {\cal S}^{\eta}_{\beta}
 ={\cal S}^{\zeta+\eta}_{\alpha+\beta}
 \end{align}
 that yields the bi-additive ${\cal S}^{\zeta+\eta}_{\alpha+\beta}$
 of two spinors ${\cal S}^{\zeta}_{\alpha}$ and ${\cal S}^{\eta}_{\beta}\in{su(2^n)}$.
 \end{defn}
 \vspace{6pt}
 With this operation, an algebraic structure in $su(2^n)$ is defined.
 \vspace{6pt}
 \begin{defn}\label{Bi-subalg}
 A set of spinors  $\mathcal{B}$ in $su(2^n)$ forms a bi-subalgebra under the bi-addition if
 ${\cal S}^{\zeta+\eta}_{\alpha+\beta}\in\mathcal{B}$
 for every pair
 ${\cal S}^{\zeta}_{\alpha}$ and
 ${\cal S}^{\eta}_{\beta}\in\mathcal{B}$.
 \end{defn}
 \vspace{6pt}
 The algebra $su(2^n)$
 is a bi-subalgebra of itself and isomorphic to the additive group $Z^{2n}_2$.
 A bi-subalgebra $\mathcal{B}$ is {\em abelian} if every pair of spinors in $\mathcal{B}$ commute,
 or {\em nonabelian} if otherwise.
 Only abelian bi-subalgebras~\cite{QAPSu1,QAPSuTsai1}
 are considered in this article.
 Refer to~\cite{QAPSuTsai2} for the examination of {\em nonabelian} bi-subalgebras.
 A maximal abelian subalgebra of $su(2^n)$,
 referred to as a {\em Cartan subalgebra} and denoted as
 $\mathfrak{C}$~\cite{QAPSu0,QAPSu1},
 is a seed to generate partitions over $su(2^n)$.
 A Cartan subalgebra $\mathfrak{C}\subset{su(2^n)}$
 is patently a bi-subalgebra isomorphic to
 $Z^n_2$ under the bi-addition.
 Being an equivalence to a subgroup of $Z^n_2$,
 a bi-subalgebra of the $k$-th maximum of $\mathfrak{C}$
 is defined~\cite{QAPSu1,QAPSuTsai1}.
 \vspace{6pt}
 \begin{defn}\label{kthMaxBiSub}
 A bi-subalgebra of the $k$-th maximum $\mathfrak{B}^{[k]}$ of a Cartan subalgebra
 $\mathfrak{C}\subset{su(2^n)}$ is a proper maximal bi-subalgebra of a bi-subalgebra of the $(k-1)$-th maximum
 $\mathfrak{B}^{[k-1]}\subset\mathfrak{C}$,
 $k=1,2,\cdots, n$.
 \end{defn}
 \vspace{6pt}
 By this definition, a Cartan subalgebra $\mathfrak{C}$ is an improper maximal
 bi-subalgebra, {\em i.e.} the {\em 0th maximal} bi-subalgebra of itself.
 A $k$-th maximal bi-subalgebra $\mathfrak{B}^{[k]}$ of $\mathfrak{C}$
 is also an $(n+k)$-th maximal bi-subalgebra of $su(2^n)$.
 As will be unfolded in the proof of Lemma~\ref{BlockinBiSubPart},
 there have nonunique Cartan subalgebras as a superset of a
 bi-subalgebra.
 Significantly, a stabilizer is an abelian bi-subalgebra.
 \vspace{6pt}
 \begin{lem}\label{StabilizerBiSub}
 The stabilizer ${\cal C}$ of a stabilizer code $[n,k,\hspace{1pt}{\cal C}]$
 is a $k$-th maximal bi-subalgebra of a Cartan subalgebra
 $\mathfrak{C}\subset{su(2^n)}$,
 and a $k$-th maximal bi-subalgebra $\mathfrak{B}^{[k]}$ of $\mathfrak{C}$ corresponds
 to the stabilizer of a stabilizer code $[n,k,\hspace{1pt}{\cal C}]$.
 \end{lem}
 \vspace{2pt}
 \begin{proof}
  To prove this lemma, it needs an assertion that an abelian set of spinors
  is a $k$-th maximal bi-subalgebra $\mathfrak{B}^{[k]}$ of a Cartan subalgebra $\mathfrak{C}\subset{su(2^n)}$
  iff this set is sized $2^{n-k}$
  and is closed under the bi-addition.
  The key of affirming the assertion lies in the fact that,
  given $\mathfrak{C}$ being isomorphic to $Z^n_2$ under the bi-addition,
  a $k$-th maximal bi-subalgebra $\mathfrak{B}^{[k]}$ of $\mathfrak{C}$ is isomorphic to a
  $k$-th maximal subgroup of $Z^n_2$,
  referring to~\cite{QAPSuTsai1} for the detailed discussion.

  A stabilizer is closed under the multiplication and
  an abelian bi-subalgebra is closed under the bi-addition.
  Since
  the multiplication of two spinors
  $(-i)^{\zeta\cdot\alpha}{\cal S}^{\zeta}_{\alpha}\cdot(-i)^{\eta\cdot\beta}{\cal S}^{\eta}_{\beta}
  =(-1)^{\eta\cdot\alpha}(-i)^{\zeta\cdot\alpha+\eta\cdot\beta}\hspace{2pt}{\cal S}^{\zeta+\eta}_{\alpha+\beta}$
  differs from their bi-additive simply by a phase,
  the isomorphism ${\cal C}=\mathfrak{B}^{[k]}$ between the two structures
  of a same size is established.
  Specifically, the isomorphism transformation,
  from the vector space of algebraic generators to the manifold of group actions,
  is the aforesaid exponential mapping
  $i(-i)^{\zeta\cdot\alpha}{\cal S}^{\zeta}_{\alpha}=e^{i\frac{\pi}{2}(-i)^{\zeta\cdot\alpha}{\cal S}^{\zeta}_{\alpha}}$.
  That is, a spinor serves both the roles of
  an algebraic generator and a group action.
\end{proof}
 \vspace{6pt}

 The QAP structure is conceived in closures.
  \vspace{6pt}
 \begin{lem}\label{Bi-subalgPart}
 Every $k$-th maximal bi-subalgebra ${\cal C}$ of a Cartan subalgebra $\mathfrak{C}\subset su(2^n)$
 can generate a partition over $su(2^n)$ consisting of  a number $2^{n+k}$ of cosets of ${\cal C}$,
 where every pair of cosets ${\cal W}_{\omega}$ and ${\cal W}_{\varsigma}$,
 $\omega$ and $\varsigma\in{Z^{n+k}_2}$,
 obey the closure under the operation of bi-addition
 $\hspace{3pt}\diamond$
 \begin{align}\label{eqPreRuleCosetofC}
 {\cal W}_{\omega}\diamond{\cal W}_{\varsigma}
 ={\cal W}_{\omega+\varsigma}.
 \end{align}
 \end{lem}
 \vspace{2pt}
 \begin{proof}
 A such structure is known as a
 {\em bi-subalgebra partition}~\cite{QAPSu0,QAPSu1,QAPSuTsai1}.
 Spinors of $su(2^n)$ form an abelian group under the
 bi-addition and ${\cal C}$ is a subgroup sized $2^{n-k}$.
 Thus, there are in total a number $2^{n+k}$ of cosets
 and each of them
 ${\cal W}_{\omega}=\{ {\cal S}^{\eta+\zeta}_{\beta+\alpha} \}$,
 indexed with a binary string $\omega\in{Z^{n+k}_2}$,
 is acquired by the bi-addition of
 ${\cal S}^{\eta}_{\beta}\notin{\cal C}$
 with all
 ${\cal S}^{\zeta}_{\alpha}\in{\cal C}$.
 Initially given
 ${\cal C}\equiv{\cal W}_{\mathbf{0}}=\{ {\cal S}^{\zeta}_{\alpha} \}$,
 the 1st coset
 ${\cal W}_{\omega_1=0\cdots 001}=\{ {\cal S}^{\eta_1+\zeta}_{\beta_1+\alpha} \}$
 is produced via the bi-addition with a spinor
 ${\cal S}^{\eta_1}_{\beta_1}\notin{\cal W}_{\mathbf{0}}$.
 The 2nd coset
 ${\cal W}_{\omega_2=0\cdots 010}=\{ {\cal S}^{\eta_2+\zeta}_{\beta_2+\alpha} \}$
 is similarly generated with another
 ${\cal S}^{\eta_2}_{\beta_2}\notin{\cal W}_{\mathbf{0}}\cup{\cal W}_{\omega_1}$.
 It follows that
 the 3rd
 ${\cal W}_{\omega_3=0\cdots 011}=\{ {\cal S}^{\eta_3+\zeta}_{\beta_3+\alpha} \}$
 is determined with the bi-addition of the two preceding cosets,
 {\em i.e.}, ${\cal S}^{\eta_3}_{\beta_3}={\cal S}^{\eta_1+\eta_2}_{\beta_1+\beta_2}$.
 Obviously, these 4 cosets are related in accord with the closure of Eq.~\ref{eqPreRuleCosetofC}.

 Let the process move on.
 By constructing the $4$th coset
 ${\cal W}_{\omega_4=0\cdots 0100}=\{ {\cal S}^{\eta_4+\zeta}_{\beta_4+\alpha} \}$
 with
 ${\cal S}^{\eta_4}_{\beta_4}\notin\bigcup^3_{m=0}{\cal W}_{\omega_m}$,
 another three cosets
 ${\cal W}_{\omega_5=0\cdots 0101}=\{ {\cal S}^{\eta_5+\zeta}_{\beta_5+\alpha} \}$,
 ${\cal W}_{\omega_6=0\cdots 0110}=\{ {\cal S}^{\eta_6+\zeta}_{\beta_6+\alpha} \}$
 and
 ${\cal W}_{\omega_7=0\cdots 0111}=\{ {\cal S}^{\eta_7+\zeta}_{\beta_7+\alpha} \}$
 are further determined,
 with the assignment
 ${\cal S}^{\eta_{\hat{m}}}_{\beta_{\hat{m}}}={\cal S}^{\eta_m+\eta_4}_{\beta_m+\beta_4}$
 and
 $\omega_{\hat{m}}=\omega_m+\omega_4$ for $0\leq m\leq 3$ and $\hat{m}=m+4$.
 This process is the so called {\em predecision rule}~\cite{QAPSu0,QAPSu1,QAPSuTsai1},
 that is, given a number $2^{p-1}$ of cosets
 ${\cal W}_{\omega_m}$, initiated from ${\cal C}={\cal W}_{\mathbf{0}}$,
 the same number of new cosets are determined through the bi-addition
 ${\cal S}^{\eta_q}_{\beta_q}\diamond{\cal W}_{\omega_m}$
 with
 ${\cal S}^{\eta_q}_{\beta_q}\notin\bigcup^{2^{p-1}-1}_{m=0}{\cal W}_{\omega_m}$,
 $0\leq m< 2^{p-1}$ and $q=2^{p-1}$.

 Assume that the first $2^{p-1}$ cosets
 ${\cal W}_{\omega_m}=\{ {\cal S}^{\eta_m+\zeta}_{\beta_m+\alpha} \}$
 satisfy the closure of Eq.~\ref{eqPreRuleCosetofC}.
 Without loss of generality, let each of them be indexed with
 a concatenation $\omega_m=\mathbf{0}\circ\upsilon_m$ of
 the $(n+k-p)$-digit string of all zeros $\mathbf{0}$ and a $p$-digit string
 $\upsilon_m=\epsilon_{m1}\cdots\epsilon_{mp}$,
  $\epsilon_{ml}\in{Z_2}$, $\epsilon_{m1}=0$,
 $0\leq m<2^{p-1}$ and $1\leq l\leq p<n+k$.
 With a spinor ${\cal S}^{\eta_q}_{\beta_q}\notin\bigcup^{2^{p-1}-1}_{m=0}{\cal W}_m$
 and thus the coset
 ${\cal W}_{\omega_q}=\{ {\cal S}^{\eta_q+\zeta}_{\beta_q+\alpha} \}\neq{\cal W}_{\omega_m}$,
 $\omega_q=\mathbf{0}\circ\upsilon_q$,
 $\upsilon_q=10\cdots 0\in{Z^p_2}$ and $q=2^{p-1}$,
 there determine a number $2^{p-1}$ of new cosets
 ${\cal W}_{\omega_{\hat{m}}}=\{ {\cal S}^{\eta_{\hat{m}}+\zeta}_{\beta_{\hat{m}}+\alpha} \}$
 via the predecision rule,
 employing
 $\omega_{\hat{m}}=\omega_m+\omega_q$,
 ${\cal S}^{\eta_{\hat{m}}}_{\beta_{\hat{m}}}={\cal S}^{\eta_m+\eta_q}_{\beta_m+\beta_q}$
 and $\hat{m}=m+q$.
 As a result of the predicision rule, apparently every triplet ${\cal W}_{\omega_m}$,
 ${\cal W}_{\omega_q}$ and ${\cal W}_{\omega_{\hat{m}}}$
 are related by
 Eq.~\ref{eqPreRuleCosetofC}.
 Another triplet
 ${\cal W}_{\omega_{\hat{m}}}=\{ {\cal S}^{\eta_{\hat{m}}+\zeta}_{\beta_{\hat{m}}+\alpha} \}$,
 ${\cal W}_{\omega_{\hat{l}}}=\{ {\cal S}^{\eta_{\hat{l}}+\bar{\zeta}}_{\beta_{\hat{l}}+\bar{\alpha}} \}$
 and ${\cal W}_{\omega_{\hat{m}}+\omega_{\hat{l}}}$
 also comply with Eq.~\ref{eqPreRuleCosetofC}
 because of the inclusion
 ${\cal S}^{\eta_{\hat{m}}+\zeta+\eta_{\hat{l}}+\bar{\zeta}}_{\beta_{\hat{m}}+\alpha+\beta_{\hat{l}}+\bar{\alpha}}
 ={\cal S}^{\eta_m+\eta_l+\zeta+\bar{\zeta}}_{\beta_m+\beta_l+\alpha+\bar{\alpha}}\in{\cal W}_{\omega_{\hat{m}}+\omega_{\hat{l}}}$,
 here $\hat{m}=m+q$, $\hat{l}=l+q$, $0\leq m,l<2^{p-1}$, and
 $\omega_{\hat{m}}+\omega_{\hat{l}}=(\omega_m+\omega_q)+(\omega_l+\omega_q)=\omega_m+\omega_l$.
 Likewise, since the bi-additive of two spinors from
 ${\cal W}_{\omega_m}$ and ${\cal W}_{\omega_{\hat{l}}}$ belongs to
 ${\cal W}_{\omega_m+\omega_{\hat{l}}}$ indexed with
 $\omega_m+\omega_{\hat{l}}=\omega_m+(\omega_l+\omega_q)$,
 these three cosets obey Eq.~\ref{eqPreRuleCosetofC}.
 Therefore, the closure of Eq.~\ref{eqPreRuleCosetofC} is fulfilled for all the $2^p$ cosets.
 \end{proof}
 \vspace{6pt}
 There comes forth the notion of syndrome.
\vspace{6pt}
\begin{defn}\label{defsyndrome}
 Given an ordered set of a number $n-k$ of independent spinors
 $\mathbf{S}_{{\cal C}}=\{{\cal S}^{\zeta_r}_{\alpha_r}:r=1,2,\cdots,n-k\}$ chosen from a bi-subalgebra ${\cal C}\subset{su(2^n)}$,
 the syndrome of a spinor ${\cal S}^{\eta}_{\beta}\in{su(2^n)}$
 with respect to $\mathbf{S}_{{\cal C}}$
 is an $(n-k)$-digit binary string
 $\tau=\epsilon_1\epsilon_2\cdots\epsilon_{n-k}$ orderly concatenated by the parities
 $\epsilon_r=\eta\cdot\alpha_r+\zeta_{r}\cdot\beta\in{Z_2}$.
\end{defn}
\vspace{6pt}
 The {\em independence} among spinors is referred to as being
 independent under the bi-addition.
 And, when the {\em syndrome} of a spinor is addressed,
 an ordered set $\mathbf{S}_{{\cal C}}$
 of detection operators is assumed.
 By tagging the feature of syndrome on each spinor in a
 partition,
 a refined version of closure emerges.
 \vspace{6pt}
 \begin{lem}\label{BlockinBiSubPart}
  The partition generated by a $k$-th maximal bi-subalgebra ${\cal C}$
  of a Cartan subalgebra in $su(2^n)$ groups into a number
  $2^{n-k}$ of blocks,
  each of which consists of $2^{2k}$ cosets of a same syndrome,
  and the closure under the bi-addition
  $\hspace{3pt}\diamond$
  \begin{align}\label{CosetCloseBiadd}
 {\cal W}_{\tau,\hspace{1pt}\mu}\diamond{\cal W}_{\upsilon,\hspace{1pt}\nu}
 ={\cal W}_{\tau+\upsilon,\hspace{1pt}\mu+\nu}
 \end{align}
 holds for every pair of cosets
 ${\cal W}_{\tau,\hspace{1pt}\mu}$
 and ${\cal W}_{\upsilon,\hspace{1pt}\nu}$
 in blocks $\Gamma_{\tau}=\bigcup_{\mu\in{Z^{2k}_2}}{\cal W}_{\tau,\hspace{1pt}\mu}$
 and $\Gamma_{\upsilon}=\bigcup_{\nu\in{Z^{2k}_2}}{\cal W}_{\upsilon,\hspace{1pt}\nu}$,
 $\tau,\upsilon\in{Z^{n-k}_2}$.
 \end{lem}
 \vspace{2pt}
 \begin{proof}
 Given an ordered set $\mathbf{S}_{{\cal C}}$
 of $n-k$ independent members taken from ${\cal C}$
 by Definition~\ref{defsyndrome},
 there are in total $2^{n+k}$ spinors
 sharing a same syndrome $\tau\in{Z^{n+k}_2}$,
 against $\mathbf{S}_{{\cal C}}$,
 that form a block $\Gamma_{\tau}$.
 Since all $2^{n-k}$ spinors of a coset of ${\cal C}$ exhibit an identical syndrome,
 a block $\Gamma_{\tau}=\bigcup_{\mu\in{Z^{2k}_2}}{\cal W}_{\tau,\hspace{1pt}\mu}$
 comprises a number $2^{2k}$ of cosets.
 Each coset ${\cal W}_{\tau,\hspace{1pt}\mu}$
 is subscripted with a block index  $\tau\in{Z^{n-k}_2}$ and a coset index
 $\mu\in{Z^{2k}_2}$.
 The single subscript $\omega\in{Z^{2k}_2}$
 of a coset ${\cal W}_{\omega}$ of Eq.~\ref{eqPreRuleCosetofC}
 then can be optionally written as the concatenation
 $\omega=\tau\circ\mu$.

 The cosets of the {\em seed block}
 $\Gamma_{\mathbf{0}}=\bigcup_{\mu\in{Z^{2k}_2}}{\cal W}_{\mathbf{0},\hspace{1pt}\mu}$
 of syndrome $\mathbf{0}$
 conform with the relation of Eq.~\ref{CosetCloseBiadd} owing to the closure
 of Eq.~\ref{eqPreRuleCosetofC} within this block, {\em i.e.},
 ${\cal W}_{\mathbf{0},\hspace{1pt}\mu}\diamond{\cal W}_{\mathbf{0},\hspace{1pt}\nu}
 ={\cal W}_{\mathbf{0},\hspace{1pt}\mu+\nu}$.
 Resorting to the predecision rule,
 a number $2k$ of spinors, respectively chosen from the same number of independent cosets in
 $\Gamma_{\mathbf{0}}-{\cal C}$,
 span a Lie algebra isomorphic to $su(2^k)$,
 referring to~\cite{QAPSu0,QAPSu1}
 and the proof of Lemma~\ref{BlocksIntrinsic-n-k}.
 Spinors of this algebra are often regarded as the {\em elementary operation}.

 A block
 $\Gamma_{\tau}=S_{\tau,\mathbf{0}}\diamond\Gamma_{\mathbf{0}}$
 other than $\Gamma_{\mathbf{0}}$ is bred through the bi-addition
 of every spinor in $\Gamma_{\mathbf{0}}$
 with $S_{\tau,\mathbf{0}}\in{su(2^n)-\Gamma_{\mathbf{0}}}$.
 That is,
 ${\cal W}_{\tau,\hspace{1pt}\mu}=S_{\tau,\mathbf{0}} \diamond {\cal W}_{\mathbf{0},\hspace{1pt}\mu}$
 for each coset ${\cal W}_{\mathbf{0},\hspace{1pt}\mu}\subset\Gamma_{\mathbf{0}}$.
 As a result, the bi-addition
 of two cosets
 ${\cal W}_{\tau,\hspace{1pt}\mu}=S_{\tau,\mathbf{0}}\diamond{\cal W}_{\mathbf{0},\hspace{1pt}\mu}$
 and
 ${\cal W}_{\upsilon,\hspace{1pt}\nu}=S_{\upsilon,\mathbf{0}}\diamond{\cal W}_{\mathbf{0},\hspace{1pt}\nu}$
 equals the coset
 $S_{\tau+\upsilon,\mathbf{0}}\diamond{\cal W}_{\mathbf{0},\hspace{1pt}\mu+\nu}={\cal W}_{\tau+\upsilon,\hspace{1pt}\mu+\nu}$,
 $S_{\tau+\upsilon,\mathbf{0}}\in{\cal W}_{\tau+\upsilon,\hspace{1pt}\mathbf{0}}$.
 This affirms Eq.~\ref{CosetCloseBiadd}.

  Remark supplementarily to Lemma~\ref{Bi-subalgPart} on nonunique Cartan subalgebras as a
  superset of an abelian bi-subalgebra.
  In the partition generated by an $(n+k)$-th maximal bi-subalgebra $\mathfrak{B}^{[k]}={\cal C}$ of $su(2^n)$
  under the bi-addition,
  there exist a number $2^{2k}$ of cosets ${\cal W}_{\mathbf{0},\hspace{1pt}\mu}\subset\Gamma_{\mathbf{0}}$
  commuting with ${\cal C}$, $\mu\in{Z^{2k}_2}$.
  Thus, there appear a number $2^{2k}-1$ of $(n+k-1)$-th maximal bi-subalgebras
  ${\cal C}\cup{\cal W}_{\mathbf{0},\hspace{1pt}\mu\neq\mathbf{0}}$
  as a superset of ${\cal C}$.
  Recursively,
  there find at the $t$-th step a number $\prod^t_{u=1}\frac{2^{2k-u-1}-1}{2^u-1}$
  of $(n+k-t)$-th maximal bi-subalgebras
  including ${\cal C}$, $1\leq t\leq k$.
  Finally as $t=k$,
  it attains in total a number
  $\prod^k_{u=1}\frac{2^{2k-u-1}-1}{2^u-1}=\prod^k_{s=1}(2^{s}-1)$ of Cartan subalgebras being a superset of ${\cal C}$,
  referring to~\cite{QAPSuTsai1} for the detailed proof.
 \end{proof}
\vspace{6pt}
 \vspace{6pt}
 \begin{thm}\label{QAPoversu2^n}
  The Lie algebra $su(2^n)$ admits the structure of
  quotient algebra partition
  $\{ \mathcal{P}_{{\cal Q}}({\cal C}) \}
  \\
  =\{ W^{\epsilon}_{\tau,\hspace{1pt}\mu}:\tau\in Z^{n-k}_2,\mu\in{Z^{2k}_2},\epsilon\in{Z_2}\}$,
  generated by a $k$-th maximal bi-subalgebra ${\cal C}$ of a Cartan
  subalgebra,
  consisting of conditioned subspaces $W^{\epsilon}_{\tau,\hspace{1pt}\mu}$
  governed with the closure
  $[W^{\epsilon}_{\tau,\hspace{1pt}\mu},W^{\sigma}_{\upsilon,\hspace{1pt}\nu}]\subset
 W^{\epsilon+\sigma}_{\tau+\upsilon,\hspace{1pt}\mu+\nu}$,
 $\upsilon\in{Z^{n-k}_2}$, $\nu\in{Z^{2k}_2}$ and
 $\sigma\in{Z_2}$.
 \end{thm}
 \vspace{2pt}
 \begin{proof}
 In the partition generated by ${\cal C}$,
 every given ${\cal S}^{\xi}_{\gamma}\in{\cal W}_{\tau,\hspace{1pt}\mu}\subset\Gamma_{\tau\neq \mathbf{0}}$
 commutes with a half of spinors ${\cal B}=\{ {\cal S}^{\zeta}_{\alpha}\}$ of ${\cal C}$
 and anticommutes with the other half
 ${\cal B}^c={\cal C}-{\cal B}=\{ {\cal S}^{\bar{\zeta}}_{\bar{\alpha}}\}$
 owing to the inclusions
 ${\cal S}^{\zeta+\eta}_{\alpha+\beta}\in\mathcal{B}$
 and  ${\cal S}^{\bar{\zeta}+\bar{\eta}}_{\bar{\alpha}+\bar{\beta}}\in\mathcal{B}$
 for all ${\cal S}^{\zeta}_{\alpha},{\cal S}^{\eta}_{\beta}\in\mathcal{B}$
 and
 ${\cal S}^{\bar{\zeta}}_{\bar{\alpha}},{\cal S}^{\bar{\eta}}_{\bar{\beta}}\in\mathcal{B}^c$~\cite{QAPSu0,QAPSu1}.
 Accordingly, each coset ${\cal W}_{\tau,\hspace{1pt}\mu}$
 is bisected into two abelian subspaces
 $W^{\epsilon}_{\tau,\mu}=\{{\cal S}^{\xi+\zeta}_{\gamma+\alpha}\}$
 and
 $W^{1+\epsilon}_{\tau,\mu}=\{{\cal S}^{\xi+\bar{\zeta}}_{\gamma+\bar{\alpha}}\}$,
 respectively called a {\em conditioned subspace}~\cite{QAPSu0,QAPSu1,QAPSuTsai1}.
 These two subspaces anticommute,
 namely $[{\cal S}^{\xi+\zeta}_{\gamma+\alpha},{\cal S}^{\xi+\bar{\zeta}}_{\gamma+\bar{\alpha}}]\neq 0$
 for every pair ${\cal S}^{\xi+\zeta}_{\gamma+\alpha}\in W^\epsilon_{\tau,\mu}$
 and ${\cal S}^{\xi+\bar{\zeta}}_{\gamma+\bar{\alpha}}\in W^{1+\epsilon}_{\tau,\mu}$.
 Whilst, since every spinor in the seed block $\Gamma_{\mathbf{0}}$ commutes with ${\cal C}$,
 it obtains $\mathcal{B}={\cal C}$ and $\mathcal{B}^c=\{0\}$.
 Each coset ${\cal W}_{\mathbf{0},\mu}\subset\Gamma_{\mathbf{0}}$
 is thus divided into a non-null subspace
 $W^{\epsilon}_{\mathbf{0},\mu}={\cal W}_{\mathbf{0},\mu}$
 and a null $W^{1+\epsilon}_{\mathbf{0},\mu}=\{0\}$.
 Reserving null subspaces in the QAP structure is essential to exhaustively generating
 $\mathfrak{t}$-$\mathfrak{p}$ decompositions
 and Cartan decompositions over $su(2^n)$~\cite{QAPSu0,QAPSu1,QAPSuTsai1}.

 Consider three coset-pairs of non-null conditioned subspaces
 $\{W^{\epsilon}_{\tau,\hspace{1pt}\mu},\hspace{2pt} W^{1+\epsilon}_{\tau,\hspace{1pt}\mu}\}$,
 $\{W^{\sigma}_{\upsilon,\hspace{1pt}\nu},\hspace{2pt} W^{1+\sigma}_{\upsilon,\hspace{1pt}\nu}\}$
 and
 $\{W^{1+\epsilon+\sigma}_{\tau+\upsilon,\hspace{1pt}\mu+\nu},\hspace{2pt} W^{\epsilon+\sigma}_{\tau+\upsilon,\hspace{1pt}\mu+\nu}\}$.
 Assume two noncommuting subspaces
 $W^{\epsilon}_{\tau,\hspace{1pt}\mu}$ and $W^{\sigma}_{\upsilon,\hspace{1pt}\nu}$
 and four spinors
 ${\cal S}^{\xi}_{\gamma},{\cal S}^{\xi+\zeta}_{\gamma+\alpha}\in W^{\epsilon}_{\tau,\hspace{1pt}\mu}$
 and
 ${\cal S}^{\pi}_{\kappa},{\cal S}^{\pi+\eta}_{\kappa+\beta}\in
 W^{\sigma}_{\upsilon,\hspace{1pt}\nu}$
 with
 ${\cal S}^{\zeta}_{\alpha}$ and ${\cal S}^{\eta}_{\beta}\in{\cal C}$,
 subject to $[{\cal S}^{\xi}_{\gamma},{\cal S}^{\pi}_{\kappa}]\neq 0$
 and
 $[{\cal S}^{\xi+\zeta}_{\gamma+\alpha},{\cal S}^{\pi+\eta}_{\kappa+\beta}]\neq 0$.
 Let the two commuting bi-additives
 ${\cal S}^{\xi+\pi}_{\gamma+\kappa}$
 and ${\cal S}^{\xi+\pi+\zeta+\eta}_{\gamma+\kappa+\alpha+\beta}$
 be both embraced in
 $W^{\epsilon+\sigma}_{\tau+\upsilon,\hspace{1pt}\mu+\nu}$,
 {\em cf.} Lemma~\ref{BlockinBiSubPart}.
 It is coerced that the vanishing commutator of every commuting pair
 ${\cal S}^{\xi}_{\gamma}\in W^{\epsilon}_{\tau,\hspace{1pt}\mu}$
 and ${\cal S}^{\hspace{1pt}\iota}_{\omega}\in W^{\sigma}_{\upsilon,\hspace{1pt}\nu}$
 as well belongs to $W^{\epsilon+\sigma}_{\tau+\upsilon,\hspace{1pt}\mu+\nu}$.
 Then, the commutator of
 $W^{\epsilon}_{\tau,\hspace{1pt}\mu}$ and $W^{1+\sigma}_{\upsilon,\hspace{1pt}\nu}$
 must be in the subspace $W^{1+\epsilon+\sigma}_{\tau+\upsilon,\hspace{1pt}\mu+\nu}$
 due to the anticommuting of
 $W^{\sigma}_{\upsilon,\hspace{1pt}\nu}$ and $ W^{1+\sigma}_{\upsilon,\hspace{1pt}\nu}$.
 The closure
 $[W^{\epsilon}_{\tau,\hspace{1pt}\mu},W^{\sigma}_{\upsilon,\hspace{1pt}\nu}]\subset
 W^{\epsilon+\sigma}_{\tau+\upsilon,\hspace{1pt}\mu+\nu}$
 of QAP $\{ \mathcal{P}_{{\cal Q}}({\cal C}) \}$ is thence established.
 Particularly notice that this structure exists in every Lie algebra $su(N)$,
 the dimension $N$ not being restricted to a power of $2$~\cite{QAPSu0,QAPSu1,QAPSuTsai1}.
  \end{proof}
 \vspace{6pt}
 In brief, a QAP~\cite{QAPSu0,QAPSu1,QAPSuTsai1} is a partition over $su(N)$
 consisting of conditioned subspaces, descendants of cosets divided,
 closed under the commutation relation as above, referring to Fig.~\ref{FigQAPstructure-n,k}.
 This closure is further generalized to the operation of
 {\em tri-addition}~\cite{QAPSuTsai1}
 that enables algorithmic and exhaustive productions of algebraic decompositions
 and resulted factorizations of group actions.
 Catching these decompositions and factorizations is essential to optimized architecture devisings of quantum computing.
 Of significance is the universality of the QAP structure to all Lie algebras~\cite{QAPSuTsai2}.
\begin{figure}[htbp]
 \begin{center}
 \[\begin{array}{ccc}
 &
 \hspace{0pt}
 \begin{array}{c}
 {\cal C}=W^1_{\tau=\mathbf{0},\hspace{1pt}\mu=\mathbf{0}}
 \end{array}
 &
 \\
 \hspace{0pt}
 \begin{array}{c}
 \vdots
 \end{array}
 &
 &
 \begin{array}{c}
 \vdots
 \end{array}\\
 \hspace{0pt}
 \begin{array}{c}
 W^{\epsilon}_{\tau,\hspace{1pt}\mu}
 \end{array}
 &
 &
 \begin{array}{c}
 W^{1+\epsilon}_{\tau,\hspace{1pt}\mu}
 \end{array}\\
 \begin{array}{c}
 \vdots
 \end{array}
 &
 &
 \begin{array}{c}
 \vdots
 \end{array}\\
 \hspace{0pt}
 \begin{array}{c}
 W^{\sigma}_{\upsilon,\hspace{1pt}\nu}
 \end{array}
 &
 &
 \begin{array}{c}
 W^{1+\sigma}_{\upsilon,\hspace{1pt}\nu}
 \end{array}\\
 \begin{array}{c}
 \vdots
 \end{array}
 &
 &
 \begin{array}{c}
 \vdots
 \end{array}\\
 \hspace{20pt}
 \begin{array}{c}
 W^{1+\epsilon+\sigma}_{\tau+\upsilon,\hspace{1pt}\mu+\nu}\\
 \end{array}
 &
 &
 \hspace{18pt}
 \begin{array}{c}
 W^{\epsilon+\sigma}_{\tau+\upsilon,\hspace{1pt}\mu+\nu}
 \end{array}\\
 \begin{array}{c}
 \vdots
 \end{array}
 &
 &
 \begin{array}{c}
 \vdots
 \end{array}\\
 \end{array}\]
\end{center}
 \fcaption{ {The QAP generated by a $k$-th maximal bi-subalgebra ${\cal C}$ of a Cartan subalgebra in
 $su(2^n)$,
 $\tau,\upsilon\in{Z^{n-k}_2}$, $\mu,\nu\in{Z^k_2}$ and $\epsilon,\sigma\in{Z_2}$,
 consisting of conditioned subspaces closed under the commutation
 relation of Theorem~\ref{QAPoversu2^n} and the tri-addition~\cite{QAPSu0,QAPSu1,QAPSuTsai1},
 ${\cal C}$ being paired with a null subspace
 $W^0_{\mathbf{0},\hspace{1pt}\mathbf{0}}=\{0\}$. }
 \label{FigQAPstructure-n,k}}
 \end{figure}

 \vspace{6pt}
 \begin{cor}\label{StabCodeisQAP}
 A stabilizer code denoted as $[n,k,\hspace{1pt}{\cal C}]=\{ \mathcal{P}_{{\cal Q}}({\cal C}) \}$ is a QAP
 generated by a $k$-th maximal bi-subalgebra ${\cal C}$
 of a Cartan subalgebra in $su(2^n)$.
 \end{cor}
 \vspace{2pt}
 \begin{proof}
  On grounds of the exponential mapping
  $i(-i)^{\zeta\cdot\alpha}{\cal S}^{\zeta}_{\alpha}=e^{i\frac{\pi}{2}(-i)^{\zeta\cdot\alpha}{\cal S}^{\zeta}_{\alpha} }$,
  each algebraic generator ${\cal S}^{\zeta}_{\alpha}$ of $su(2^n)$
  is transformed into a group action of the same form in $SU(2^n)$.
  Through this isomorphism,
  the stabilized code
  $[n,k,\hspace{1pt}{\cal C}]$
  inherits the structure $\{ \mathcal{P}_{{\cal Q}}({\cal C}) \}$ as the partition required.
  An abelian subspace in this partition consequently converts into
  an abelian subset of the code,
  but legitimately still named an abelian subspace.
 \end{proof}
 \vspace{6pt}
  In the following, a stabilized code $[n,k,\hspace{1pt}{\cal C}]$ is thus called a partition
  and is reckoned a {\em quantum extension} of a Hamming code $[n,k]$.
  Explicitly,
  each coset $C_{\tau}$ of $[n,k]$ {\em grows} into a block $\Gamma_{\tau}$ in $[n,k,\hspace{1pt}{\cal C}]$,
  and every element of $C_{\tau}$,
  a binary string of length $n$, {\em expands} to a coset of $\Gamma_{\tau}$, $\tau\in{Z^{n-k}_2}$,
  {\em cf.} Lemma~\ref{HammingEmbed}.

 A great importance is the duality of partitions of operators and states.
  \vspace{6pt}
\begin{thm}\label{fundbasiscodewords}
 Given an ordered set of a number $n-k$ of independent spinors
 $\mathbf{S}_{{\cal C}}=\{S_r:r=1,2,\cdots,n-k\}$
 from a $k$-th maximal bi-subalgebra ${\cal C}$ of a Cartan subalgebra
 in $su(2^n)$,
 the space of $n$-qubit states admits a decomposition
 ${\cal H}=\bigoplus_{\tau\in{Z^{n-k}_2}}{\cal H}_{\tau}$,
  where each ${\cal H}_{\tau}$,
 formed by states $\ket{\psi}$ satisfying
 $S_r\ket{\psi}=(-1)^{\epsilon_r}\ket{\psi}$,
 is an eigen-invariant subspace of ${\cal C}$
 with the syndrome $\tau=\epsilon_1\epsilon_2\cdots\epsilon_{n-k}$,
 $\epsilon_r\in{Z_2}$,
 and also an invariant subspace of the seed block $\Gamma_{\mathbf{0}}$,
 i.e.,
 $\Gamma_{\mathbf{0}}({\cal H}_{\tau})\subset{\cal H}_{\tau}$;
 moreover, the duality
 $\Gamma_{\upsilon}({\cal H}_{\tau})\subset{\cal H}_{\tau+\upsilon}$
 holds
 for every block $\Gamma_{\upsilon}$
 and subspace ${\cal H}_{\tau}$, $\upsilon\in{Z^{n-k}_2}$.
\end{thm}
\vspace{2pt}
\begin{proof}
 Since spinors in ${\cal C}$ are pairwise commuting and individually diagonalizable with eigenvaules $+1$ and $-1$,
 the space of $n$-qubit states
 admits a spectral decomposition ${\cal H}=\bigoplus_{\lambda}{\cal H}_{\lambda}$
 of disjoint eigenspaces ${\cal H}_{\lambda}$
 of a unique eigenspectrum,
 {\em cf.} the spectral theorem~\cite{ZimmerSpectral}.
 That is,
 an eigenspace ${\cal H}_{\lambda}$ of ${\cal C}$
 with eigenspectrum $\lambda=(\lambda_1,\cdots,\lambda_{n-k})$
 consists of states $\ket{\psi}$ satisfying
 $S_r\ket{\psi}=\lambda_r\ket{\psi}$,
 $\lambda_r=\pm 1$,
 here $S_r$ being the $r$-th member in the ordered set $\mathbf{S}_{{\cal C}}$,
 $1\leq r\leq n-k$.
 As shown immediately that every eigenspectrum $\lambda$ corresponds to a syndrome
 $\tau\in{Z^{n-k}_2}$,
 each ${\cal H}_{\lambda}$ is referred to as an eigen-invariant subspace of ${\cal C}$
 with syndrome $\tau$.

 Accordingly, states in obedience to the constraint of $n-k$ linearly
 independent equations
 $S_r\ket{\psi}=\ket{\psi}$,
 $S_r\in \mathbf{S}_{{\cal C}}$,
 constitute a $2^k$-dimensional eigen-invariant subspace
 ${\cal H}_{\mathbf{0}}$ of ${\cal C}$ with eigenvalues $\lambda_r=+1=(-1)^{\epsilon_r=0}$,
 {\em i.e.}, of syndrome $\tau=\epsilon_1\cdots\epsilon_{n-k}=\mathbf{0}$.
 Then, there exists an orthonormal frame
 $\{\ket{\psi_{\mathbf{0},i}}:i\in{Z^k_2}\}$
 in ${\cal H}_{\mathbf{0}}$
 composed of simultaneous eigenvectors of $S_r$.
 Let each {\em axial vector} of the frame
 be a {\em basis codeword}
 and, of the syndrome $\mathbf{0}$,
 every vector in ${\cal H}_{\mathbf{0}}$ is a {\em codeword}, too.
 Thereby ${\cal H}_{\mathbf{0}}$ is regarded as the {\em codeword subspace}.
 Since, for every $S_{\mathbf{0}}\in\Gamma_{\mathbf{0}}$
 and $\ket{\psi_{\mathbf{0}}}\in{\cal H}_{\mathbf{0}}$,
 $S_r(S_{\mathbf{0}}\ket{\psi_{\mathbf{0}}})=S_{\mathbf{0}}S_r\ket{\psi_{\mathbf{0}}}=S_{\mathbf{0}}\ket{\psi_{\mathbf{0}}}$
 for all $S_r\in\mathbf{S}_{{\cal C}}$ owing to the commuting of $\Gamma_{\mathbf{0}}$ and ${\cal C}$,
 {\em cf.} Lemma~\ref{BlockinBiSubPart},
 the inclusion $S_{\mathbf{0}}\ket{\psi_{\mathbf{0}}}\in{\cal H}_{\mathbf{0}}$ is acquired,
 leading to the invariance
 $\Gamma_{\mathbf{0}}({\cal H}_{\mathbf{0}})\subset{\cal H}_{\mathbf{0}}$.

 By applying an arbitrary spinor $S_{\tau}\in\Gamma_{\tau\neq\mathbf{0}}$
 to each basis codeword $\ket{\psi_{\mathbf{0},i}}$,
 an orthonormal frame
 $\{\ket{\psi_{\tau,i}}=S_{\tau}\ket{\psi_{\mathbf{0},i}}\}$
 of dimension $2^k$ is built,
 $\tau\in{Z^{n-k}_2}$.
 In virtue of the fact
 $S_r\ket{\psi_{\tau,i}}=\lambda_r\ket{\psi_{\tau,i}}=(-1)^{\epsilon_r}\ket{\psi_{\tau,i}}$
 based on $S_rS_{\tau}=(-1)^{\epsilon_r}S_{\tau}S_r$ for every $S_r\in\mathbf{S}_{{\cal C}}$,
 each axial vector of this frame is an eigenstate of $S_r$ with eigenvalue $(-1)^{\epsilon_r}$
 and is thus considered a {\em corrupted state} of syndrome
 $\tau=\epsilon_1\epsilon_2\cdots\epsilon_{n-k}$,
 {\em cf.} Definition~\ref{defsyndrome}.
 Any state spanned by these axial vectors
 is also a corrupted state of the same syndrome.
 An eigen-invariant subspace ${\cal H}_{\tau}$
 of corrupted states with syndrome $\tau$
 is then introduced.
 Similarly, given every $S_{\mathbf{0}}\in\Gamma_{\mathbf{0}}$
 and $\ket{\psi_{\tau}}\in{\cal H}_{\tau}$,
 the eigen-relation
 $S_r(S_{\mathbf{0}}\ket{\psi_{\tau}})=S_{\mathbf{0}}S_r\ket{\psi_{\tau}}=(-1)^{\epsilon_r}S_{\mathbf{0}}\ket{\psi_{\tau}}$
 is reached
 for each $S_r\in\mathbf{S}_{{\cal C}}$,
 {\em i.e.},
 $S_{\mathbf{0}}\ket{\psi_{\tau}}\in{\cal H}_{\tau}$.
 The invariance is assured
 $\Gamma_{\mathbf{0}}({\cal H}_{\tau})\subset{\cal H}_{\tau}$.

 In general,
 for every spinor $S_{\upsilon}\in\Gamma_{\upsilon}$
 and  state $\ket{\psi_{\tau}}\in{\cal H}_{\tau}$,
 there attains
 $S_r(S_{\upsilon}\ket{\psi_{\tau}})=(-1)^{\sigma_r}S_{\upsilon}S_r\ket{\psi_{\tau}}=(-1)^{\sigma_r+\epsilon_r}S_{\upsilon}\ket{\psi_{\tau}}$
 for every $S_r\in\mathbf{S}_{{\cal C}}$
 and the $r$-th digits $\sigma_r$ and $\epsilon_r$ from $\upsilon$ and $\tau$,
  which implies $S_{\upsilon}\ket{\psi_{\tau}}\in{\cal H}_{\tau+\upsilon}$.
 This unveils the duality,
 spinors as operators vs states as operatees,
 $\Gamma_{\upsilon}({\cal H}_{\tau})\subset{\cal H}_{\tau+\upsilon}$.

 Hence, the space of $n$-qubt states
 ${\cal H}=\bigoplus_{\tau\in{Z^{n-k}_2}}{\cal H}_{\tau}$
 is partitioned into a number $2^{n-k}$ of eigen-invariant subspaces ${\cal H}_{\tau}$ of
 ${\cal C}$.
 Suppose two corrupted states
 $\ket{\psi_{\tau,i}}$ and $\ket{\psi_{\upsilon,j}}$
 are of different syndromes $\tau$
 and $\upsilon$.
 Note the existence of
 $\epsilon_{\hat{r}}\neq\sigma_{\hat{r}}$
 elicited by
 $S_{\hat{r}}\in\mathbf{S}_{{\cal C}}$,
 {\em i.e.},
 $\epsilon_{\hat{r}}$ and $\sigma_{\hat{r}}$
 being the $\hat{r}$-th digits of $\tau$ and $\upsilon$.
 The orthogonality $\braket{\psi_{\upsilon,j}}{\psi_{\tau,i}}=0$
 is rooted in the sign discrepancy
 \begin{align}\label{eqcalOrthogonal}
  \braket{\psi_{\upsilon,j}}{\psi_{\tau,i}}
 =&\bra{\psi_{\upsilon,j}}S_{\hat{r}}S_{\hat{r}}\ket{\psi_{\tau,i}}
 =(-1)^{\epsilon_{\hat{r}}+\sigma_{\hat{r}}}\braket{\psi_{\upsilon,j}}{\psi_{\tau,i}}
 =-\braket{\psi_{\upsilon,j}}{\psi_{\tau,i}}.
 \end{align}
  It concludes that eigen-invariant subspaces of distinct syndromes are orthogonal.
\end{proof}
\vspace{6pt}
 The successive is the orthogonality connecting spinors and codewords.
 \vspace{6pt}
 \begin{cor}\label{OrthogonalityCond}
 In a partition $[n,k,\hspace{1pt}{\cal C}]$,
 the orthogonality condition
 \begin{align}\label{orthocond}
  \bra{\psi_{\mathbf{0},j}}S_{\upsilon,\hspace{1pt}\nu}S_{\tau,\hspace{1pt}\mu}\ket{\psi_{\mathbf{0},i}}
  =\chi\delta_{\upsilon\tau}\delta_{ji}
  \end{align}
 holds
 for basis codewords
 $\ket{\psi_{\mathbf{0},i}}$
 and $\ket{\psi_{\mathbf{0},j}}$,
 and two spinors $S_{\tau,\hspace{1pt}\mu}\in{\cal W}_{\tau,\mu}$
 and $S_{\upsilon,\hspace{1pt}\nu}\in{\cal W}_{\upsilon,\nu}$
 either in distinct blocks
 $\Gamma_{\tau}$ and $\Gamma_{\upsilon}$, $\tau\neq\upsilon$, or in a same coset
 ${\cal W}_{\tau=\upsilon,\mu=\nu}$ of block $\Gamma_{\tau=\upsilon}$
 with
 $\chi=\pm 1$ or $\pm i$, $\tau,\upsilon\in{Z^{n-k}_2}$, $\mu,\nu\in{Z^{2k}_2}$ and $i,j\in{Z^k_2}$.
 \end{cor}
 \vspace{2pt}
 \begin{proof}
 The circumstance of two spinors in different blocks is affirmed
 by Eq.~\ref{eqcalOrthogonal}.
 Given
 $S_{\tau,\hspace{1pt}\mu}=(-i)^{(\xi+\zeta)\cdot(\gamma+\alpha)}{\cal S}^{\xi+\zeta}_{\gamma+\alpha}$
 and
 $S_{\upsilon,\hspace{1pt}\nu}=(-i)^{(\xi+\eta)\cdot(\gamma+\beta)}{\cal S}^{\xi+\eta}_{\gamma+\beta}$
 in a same coset ${\cal W}_{\tau,\mu}$
 of a block,
 ${\cal S}^{\xi}_{\gamma}\in{\cal W}_{\tau,\mu}$
 and
 ${\cal S}^{\zeta}_{\alpha},{\cal S}^{\eta}_{\beta}\in{\cal C}$,
 it derives
 $\bra{\psi_{\mathbf{0},j}}S_{\upsilon,\hspace{1pt}\nu}S_{\tau,\hspace{1pt}\mu}\ket{\psi_{\mathbf{0},i}}
 =\chi\hspace{1pt}\delta_{ij}$
 by dint of
 $S_{\upsilon,\hspace{1pt}\nu}S_{\tau,\hspace{1pt}\mu}\in{\cal C}$,
 $\chi^2=(-1)^{\xi\cdot(\alpha+\beta)+(\zeta+\eta)\cdot\gamma}=\pm 1$.
 The global phase $\chi$ differs in two scenarios,
 either $\chi=\pm 1$ if the two spinors commute,
 {\em i.e.}, both in a same abelian (conditioned) subspace,
 or $\chi=\pm i$ if they anticommute,
 respectively in one of abelian subspaces of ${\cal W}_{\tau,\hspace{1pt}\mu}$,
 {\em cf.} Theorem~\ref{QAPoversu2^n}.
 This foretells the concept of coset spinor as stated in Lemma~\ref{CosetErrorCorrectable}.
 While, the case of two spinors in distinct cosets of a block,
 errors uncorrectable,
 will be elucidated in Lemma~\ref{OrthoCondIntrisicCoord}.
 \end{proof}
 \vspace{6pt}
 The concept of {\em coset spinor} is essential throughout the work.
\vspace{6pt}
 \begin{lem}\label{CosetErrorCorrectable}
  In a partition $[n,k,\hspace{1pt}{\cal C}]$,
  the concept a spinor is a coset spinor conveys two implications,
  the correction equivalence
  that an error is correctable by any member in a same coset,
  and
  the code degeneracy that a correctable error set allows spinors in a same coset.
 \end{lem}
 \vspace{2pt}
 \begin{proof}
 To prove the first implication,
 similarly,
 assume
 $S_{\tau,\hspace{1pt}\mu}$ and
 $S'_{\tau,\hspace{1pt}\mu}$
 in a same coset
 ${\cal W}_{\tau,\hspace{1pt}\mu}$.
 Since $S'_{\tau,\hspace{1pt}\mu}S_{\tau,\hspace{1pt}\mu}\in{\cal C}$,
 the codeword
 $\chi\hspace{1pt}\ket{\psi}=S'_{\tau,\hspace{1pt}\mu}S_{\tau,\hspace{1pt}\mu}\ket{\psi}$
 is recovered from a corruption $S_{\tau,\hspace{1pt}\mu}\ket{\psi}$
 by performing the correction operator
 $S'_{\tau,\hspace{1pt}\mu}$,
 either $\chi=\pm 1$ if the two spinors commute
 or $\chi=\pm i$ if they anticommute, {\em cf.} Corollary~\ref{OrthogonalityCond}.

 The assertion that an error of
 spinor is correctable by any member of a same coset bears a {\em reciprocal}
 implication that errors of a same coset can be corrected
 by any spinor of the same coset,
 namely the code degeneracy.
 \end{proof}
\vspace{6pt}
 Not simply for herself,
 a spinor stands for every fellow member of a same coset in terms of
 the correction equivalence and the code degeneracy.

 The continued assertions concern the error correction within a partition.
  \vspace{6pt}
 \begin{lem}\label{ErrorDetectable}
  An error set ${\cal E}$ is detectable by a partition $[n,k,\hspace{1pt}{\cal C}]$
  if no spinor of ${\cal E}$ is in the subspace
  $\Gamma_{\mathbf{0}}-{\cal C}$.
 \end{lem}
 \vspace{2pt}
 \begin{proof}
  To be detectable, an error has to exhibit a syndrome
  $\tau\neq\mathbf{0}$.
 \end{proof}
\vspace{6pt}
\vspace{6pt}
 \begin{lem}\label{TwoCosetErrorUnCorrectable}
  In a partition $[n,k,\hspace{1pt}{\cal C}]$,
  two errors are uncorrectable if they are in distinct cosets of
  a same block.
 \end{lem}
 \vspace{2pt}
 \begin{proof}
  Let
  $S_{\tau,\hspace{1pt}\mu}\in{\cal W}_{\tau,\hspace{1pt}\mu}$
  and
  $S_{\tau,\hspace{1pt}\nu}\in{\cal W}_{\tau,\hspace{1pt}\nu}$
  be two such spinors,
  $\mu,\nu\in{Z^{2k}_2}$ and $\mu\neq\nu$.
  With a codeword $\ket{\psi}$ of ${\cal C}$,
  the corrupted states
  $S_{\tau,\hspace{1pt}\mu}\ket{\psi}$
  and
  $S_{\tau,\hspace{1pt}\nu}\ket{\psi}$
  are indistinguishable due to a same syndrome
  $\tau\in{Z^{n-k}_2}$.
  Since
  $S_{\tau,\hspace{1pt}\nu}S_{\tau,\hspace{1pt}\mu}\in\Gamma_{\mathbf{0}}-{\cal C}$,
  a mistaken correction
  $S_{\tau,\hspace{1pt}\nu}S_{\tau,\hspace{1pt}\mu}\ket{\psi}$
  turns to be an undetectable corrupted state,
  albeit remaining in the invariant subspace ${\cal H}_{\mathbf{0}}$.
 \end{proof}
\vspace{6pt}
 \vspace{6pt}
\begin{thm}\label{ErrorSetcorrectable}
 An error set ${\cal E}$ is correctable
 by a partition $[n,k,\hspace{1pt}{\cal C}]$
 iff two arbitrary spinors of ${\cal E}$ are either in different blocks
 or in a same coset of a block within this partition.
\end{thm}
 \vspace{2pt}
 \begin{proof}
 This theorem is a direct consequence of Lemmas~\ref{CosetErrorCorrectable}
 and~\ref{TwoCosetErrorUnCorrectable}.
 Notice that the code degeneracy is by default admitted.
 The prerequisite is placed that
 no spinor of ${\cal E}$ is allowed in $\Gamma_{\mathbf{0}}-{\cal C}$,
 {\em cf.} Lemma~\ref{ErrorDetectable}.
 \end{proof}
 \vspace{6pt}
 The condition of $t$-error correction
 is induced straightaway.
  \vspace{6pt}
\begin{cor}\label{t-errorCorrectable}
 A partition corrects all errors of weight
 $\leq t$,
 suggestively denoted as $[n,k,\hspace{1pt}{\cal C};t]$,
 iff every spinor in the subspace $\Gamma_{\mathbf{0}}-{\cal C}$ is of weight
 $\geq 2t+1$.
\end{cor}
 \vspace{2pt}
 \begin{proof}
 It is reasonable in practice to presume $t< n/2$.
 Noteworthily, every spinor of weight $\leq 2t$ is also a bi-additive of
 two errors of weight $\leq t$.
 Since two correctable errors
  by Theorem~\ref{ErrorSetcorrectable}
 are either in distinct blocks or in a same coset of a block,
 the subspace $\Gamma_{\mathbf{0}}-{\cal C}$ embraces no spinor of weight $\leq 2t$,
 which validates one implication.
 The assumption of the other implication
 compels the product, of weight $\leq 2t$, of every pair of errors  either in
 ${\cal C}$ or in a block $\Gamma_{\tau\neq\mathbf{0}}$,
 that is, the two spinors are either in a same coset of a block
 or in distinct blocks.
 This is in agreement with the condition
 of Theorem~\ref{ErrorSetcorrectable}.
 \end{proof}
 \vspace{6pt}

 \section{Intrinsic Coordinate}\label{secIntrinsicBiSub}
 To prepare fault tolerant encodes in a partition,
 a special coordinate is required.
 \vspace{6pt}
 \begin{defn}\label{defnIntCartan}
 The intrinsic Cartan subalgebra
 \begin{align}\label{eqIntCartan}
  \mathfrak{C}_{[\mathbf{0}]}
 =\{ {\cal S}^{\xi}_{\mathbf{0}}:\xi\in{Z^{n}_2} \}
 \end{align}
 is a maximal abelain subalgebra of $su(2^n)$ composed of diagonal spinors of $n$ qubits.
 \end{defn}
 \vspace{6pt}
 The set of phase strings $\{\xi\}$ from
 $\mathfrak{C}_{[\mathbf{0}]}$
 assembles the group $Z^n_2$ under the bitwise addition.

 \vspace{6pt}
 \begin{lem}\label{kthmaxinIntCartan}
 A bi-subalgebra ${\cal C}=\{{\cal S}^{\xi}_{\mathbf{0}}\}$
 is a diagonal bi-subalgebra of the $k$-th maximum of the intrinsic Cartan subalgebra $\mathfrak{C}_{[\mathbf{0}]}\subset{su(2^n)}$
 if the phase strings $\{\xi\}$ of spinors in ${\cal C}$
 form a $k$-th maximal subgroup of $Z^n_2$, $k=0,1,\cdots,n$.
 \end{lem}
 \vspace{2pt}
 \begin{proof}
  By Definition~\ref{kthMaxBiSub}, ${\cal C}$ is $k$-th maximal in $\mathfrak{C}_{[\mathbf{0}]}$.
 \end{proof}
 \vspace{6pt}
   There exist a number $\prod^{n-k}_{r=0}\frac{2^{n-r+1}-1}{2^r-1}$ of $k$-th maximal
   bi-subalgebras
   in every Cartan subalgebra $\mathfrak{C}$ of $su(2^n)$,
   including the intrinsic $\mathfrak{C}_{[\mathbf{0}]}$~\cite{QAPSu0,QAPSu1,QAPSuTsai1}.
 \vspace{6pt}
 \begin{lem}\label{HammingEmbed}
 Every Hamming code $[n,k]$ is embeddable into the partition
 $[n,k,\hspace{1pt}{\cal C}]$ generated by a $k$-th maximal bi-subalgebra
 ${\cal C}$ of a Cartan subalgebra in $ su(2^n)$.
 \end{lem}
 \vspace{2pt}
 \begin{proof}
  Without loss of generality,
  a choice of partition yielt with a diagonal $k$-th maximal bi-subalgebra
  of the intrinsic Cartan subalgebra $\mathfrak{C}_{[\mathbf{0}]}\subset{su(2^n)}$
  is here provided.
 A Hamming code $[n,k]$ is known to be a partition under the bitwise addition,
 consisting of  a number $2^{n-k}$ of cosets
 $C_{\tau}=\{\beta_{\tau}+\alpha_h\}$
 with syndrome $\tau\in{Z^{n-k}_2}$,
 $\beta_{\tau}\in{Z^n_2}$, of an $(n-k)$-th maximal subgroup
 $C_{\mathbf{0}}=\{\alpha_h:1\leq h\leq 2^k\}$ of $Z^n_2$.
 A partition $[n,k,\hspace{1pt}{\cal C}]$ feasible to the embedding
 desired
 is generated by the $k$-th maximal bi-subalgebra
 ${\cal C}=\{{\cal S}^{\xi}_{\mathbf{0}}\}\subset\mathfrak{C}_{[\mathbf{0}]}$
 satisfying
 $\xi\cdot\alpha_h=0$ for every $\alpha_h\in{C_0}$, $\xi\in{Z^n_2}$.

 A little more precisely,
  every string $\alpha_h\in{C_0}$
 expands to a coset
  ${\cal W}_{\mathbf{0},\hspace{1pt}\mu_h}
 =\{{\cal S}^{\xi}_{\alpha_h}={\cal S}^{\mathbf{0}}_{\alpha_h}\diamond{\cal S}^{\xi}_{\mathbf{0}}:{\cal S}^{\xi}_{\mathbf{0}}\in{\cal C}\}$
 of ${\cal C}$
 in the seed block $\Gamma_{\mathbf{0}}$, $\mu_h\in{Z^{2k}_2}$ and $1\leq h\leq 2^k$.
 In the intrinsic Cartan subalgebra $\mathfrak{C}_{[\mathbf{0}]}\subset\Gamma_{\mathbf{0}}$,
 there exist a number $2^k$ of cosets
 ${\cal W}_{\mathbf{0},\hspace{1pt}\nu_m}
 =\{ {\cal S}^{\theta_m+\xi}_{\mathbf{0}}={\cal S}^{\theta_m}_{\mathbf{0}}\diamond{\cal S}^{\xi}_{\mathbf{0}}:{\cal S}^{\xi}_{\mathbf{0}}\in{\cal C} \}$
 of ${\cal C}$,
 here ${\cal S}^{\theta_m}_{\mathbf{0}}\in\mathfrak{C}_{[\mathbf{0}]}$, $\nu_m\in{Z^{2k}_2}$ and $1\leq m\leq 2^k$.
  Thus, the seed coset $C_0$
  grows into the block
  $\Gamma_{\mathbf{0}}$ formed by the $2^{2k}$ cosets
  $\{  {\cal W}_{\mathbf{0},\hspace{1pt}\mu_h+\nu_m}={\cal W}_{\mathbf{0},\hspace{1pt}\mu_h}\diamond{\cal W}_{\mathbf{0},\hspace{1pt}\nu_m}\}$.
  Similarly, each string $\beta_{\tau}+\alpha_h$ of the coset $C_{\tau}$ of syndrome $\tau\neq\mathbf{0}$
  stretches to a coset
  ${\cal W}_{\tau,\hspace{1pt}\mu_h}=
  \{ {\cal S}^{\xi}_{\beta_{\tau}+\alpha_h}
  ={\cal S}^{\mathbf{0}}_{\beta_{\tau}+\alpha_h}\diamond{\cal S}^{\xi}_{\mathbf{0}}:{\cal S}^{\xi}_{\mathbf{0}}\in{\cal C} \}
  \subset\Gamma_{\tau}$,
  ${\cal S}^{\mathbf{0}}_{\beta_{\tau}+\alpha_h}\in\Gamma_{\tau}$,
  and  the coset $C_{\tau}$
  enlarges to the block
 $\Gamma_{\tau}$ of syndrome $\tau$ constituted by the $2^{2k}$ cosets
 $\{  {\cal W}_{\tau,\hspace{1pt}\mu_h+\nu_m}={\cal W}_{\tau,\hspace{1pt}\mu_h}\diamond{\cal W}_{\mathbf{0},\hspace{1pt}\nu_m}\}$,
 {\em cf.} Lemma~\ref{BlockinBiSubPart}.
  \end{proof}
 \vspace{6pt}
 The partition $[n,k,\hspace{1pt}{\cal C}]$ is a {\em quantum extension} of a Hamming code $[n,k]$.

 Amongst bi-subalgebras of $\mathfrak{C}_{[\mathbf{0}]}$,
 the {\em intrinsic} bi-subalgebra accommodates the {\em intrinsic coordinate}.
\vspace{6pt}
 \begin{lem}\label{BiSubIntrCoord}
 The intrinsic bi-subalgebra of the $k$-th maximum
 \begin{align}\label{Intrinsickthbi-Sub}
 \hat{{\cal C}}=\{ {\cal S}^{\zeta}_{\mathbf{0}}\otimes{\cal S}^{\mathbf{0}}_{\mathbf{0}}:\zeta\in{Z^{n-k}_2} \},
\end{align}
 of the intrinsic Cartan subalgebra $\mathfrak{C}_{[\mathbf{0}]}\subset{su(2^n)}$,
 has the eigenstates $\ket{\tau}\otimes\ket{i}$ that form the intrinsic coordinate,
 $\tau\in{Z^{n-k}_2}$ and $i\in Z^k_2$,
 here ${\cal S}^{\zeta}_{\mathbf{0}}$ being a diagonal spinor of ${su(2^{n-k})}$
 and
 ${\cal S}^{\mathbf{0}}_{\mathbf{0}}$ the identity of ${su(2^k)}$.
 \end{lem}
\vspace{2pt}
 \begin{proof}
 Obviously, $\hat{{\cal C}}$ is also $k$-th maximal in
 $\mathfrak{C}_{[\mathbf{0}]}$.
 Since ${\cal S}^{\zeta}_{\mathbf{0}}\otimes{\cal S}^{\mathbf{0}}_{\mathbf{0}}\ket{\tau}\otimes\ket{i}
  =(-1)^{\zeta\cdot\tau}\ket{\tau}\otimes\ket{i}$ via Eq.~\ref{spinoronstate},
  each $\ket{\tau}\otimes\ket{i}$ is an eigenstate of
  ${\cal S}^{\zeta}_{\mathbf{0}}\otimes{\cal S}^{\mathbf{0}}_{\mathbf{0}}\in\hat{{\cal C}}$
  with eigenvaule $(-1)^{\zeta\cdot\tau}$.
 The basis state $\ket{\tau}$ of $n-k$ qubits may be referred to as the
 {\em encode} part that carries a syndrome,
 and the $k$-qubit basis state $\ket{i}$ as the {\em origin} part.
 \end{proof}
 \vspace{6pt}
 Recall that,
 bridged by the exponential mapping
 $i(-i)^{\xi\cdot\gamma}{\cal S}^{\xi}_{\gamma}=e^{i\frac{\pi}{4}(-i)^{\xi\cdot\gamma}{\cal S}^{\xi}_{\gamma}}$,
 a spinor ${\cal S}^{\xi}_{\gamma}$ plays both the roles of an algebraic generator
 and a group action.
 In the intrinsic coordinate,
 the QAP structure is read in exact components.
 \vspace{6pt}
 \begin{lem}\label{BlocksIntrinsic-n-k}
 In the partition $[n,k,\hspace{1pt}\hat{{\cal C}}]$ generated by the
 intrinsic bi-subalgebra  $\hat{{\cal C}}\subset su(2^n)$,
 each block $\hat{\Gamma}_{\tau}=\bigcup_{\mu\in{Z^{2k}_2}} \hat{{\cal W}}_{\tau,\hspace{1pt}\mu}$
 consists of  a number $2^{2k}$ of cosets of the same syndrome
 $\tau\in{Z^{n-k}_2}$
 \begin{align}\label{CosetkIntrinsic}
 \hat{{\cal W}}_{\tau,\hspace{1pt}\mu}
 =\{ {\cal S}^{\zeta}_{\tau}\otimes{\cal S}^{\varsigma}_{\kappa}:\zeta\in{Z^{n-k}_2} \},
\end{align}
 subscripted with the coset index
 $\mu=\varsigma\circ\kappa\in{Z^{2k}_2}$,
 ${\cal S}^{\zeta}_{\tau}\in{su(2^{n-k})}$
 and
 ${\cal S}^{\varsigma}_{\kappa}\in{su(2^k)}$.
 \end{lem}
 \vspace{2pt}
 \begin{proof}
 This lemma is Lemma~\ref{BlockinBiSubPart} in the intrinsic coordinate.
 Remind that by Definition~\ref{defsyndrome}
 the syndrome of a coset is etched with respect to an ordered set of $n-k$
 independent spinors taken from $\hat{{\cal C}}$, for instance the set $\hat{S}_{\hat{{\cal C}}}$
 as of Lemma~\ref{IntrCoordFundaLemmaCodeword}.
 The seed block
 $\hat{\Gamma}_{\mathbf{0}}=\bigcup_{\mu\in{Z^{2k}_2}} \hat{{\cal W}}_{\mathbf{0},\hspace{1pt}\mu}$
 is the union of $2^{2k}$ cosets of the form
 $\hat{{\cal W}}_{\mathbf{0},\hspace{1pt}\mu}
 =\{ {\cal S}^{\zeta}_{\mathbf{0}}\otimes{\cal S}^{\varsigma}_{\kappa}:\zeta\in{Z^{n-k}_2} \}$,
 $\mu=\varsigma\circ\kappa\in{Z^{2k}_2}$.
 Each coset
 $\hat{{\cal W}}_{\tau,\hspace{1pt}\mu}=\{ {\cal S}^{\zeta}_{\tau}\otimes{\cal S}^{\varsigma}_{\kappa}:\zeta\in{Z^{n-k}_2} \}$
 of block $\Gamma_{\tau}$,
 pertaining to the coset index $\mu=\varsigma\circ\kappa$,
 is produced through the bi-addition of
 spinors in $\hat{{\cal W}}_{\mathbf{0},\hspace{1pt}\mu}\subset\hat{\Gamma}_{\mathbf{0}}$
 with
 ${\cal S}^{\mathbf{0}}_{\tau}\otimes{\cal S}^{\mathbf{0}}_{\mathbf{0}}\in\hat{\Gamma}_{\tau}$.
 The linking amongst cosets is evinced in terms of the closure that the bi-additive
    ${\cal S}^{\zeta+\eta}_{\tau+\upsilon}\otimes{\cal S}^{\varsigma+\pi}_{\kappa+\omega}$
    of two generators
    ${\cal S}^{\zeta}_{\tau}\otimes{\cal S}^{\varsigma}_{\kappa}\in\hat{{\cal W}}_{\tau,\hspace{1pt}\mu}$
    and
    ${\cal S}^{\eta}_{\upsilon}\otimes{\cal S}^{\pi}_{\omega}\in\hat{{\cal W}}_{\upsilon,\hspace{1pt}\nu}$
    is in
    $\hat{{\cal W}}_{\tau+\upsilon,\hspace{1pt}\mu+\nu}$,
    $\nu=\pi\circ\omega$.
    The operator ${\cal S}^{\zeta}_{\tau}$
    acts on states of $n-k$ qubits,
    and ${\cal S}^{\varsigma}_{\kappa}$
    on those of $k$ qubits.
  Furthermore, all spinors in
  $\hat{{\cal W}}_{\tau,\hspace{1pt}\mu}$
  share an identical part of $k$ qubits ${\cal S}^{\varsigma}_{\kappa}$,
  noting $\mu=\varsigma\circ\kappa$.
 It is therefore legitimate to identify the $k$-qubit part ${\cal S}^{\varsigma}_{\kappa}$
 with the coset $\hat{{\cal W}}_{\tau,\hspace{1pt}\mu}$.
 This identification is
 a manifestation of the concept of coset spinor,
 {\em cf.} Lemma~\ref{CosetErrorCorrectable},
 and is crucial to the construction of fault tolerant encodes in Section~\ref{secFTQC}.

 Similar to the procedure of Lemma~\ref{BlockinBiSubPart},
 let a number $2k$ of spinors
 $\{ {\cal S}^{\zeta_p}_{\mathbf{0}}\otimes{\cal S}^{\varsigma_p}_{\kappa_p}:1\leq p\leq 2k \}$
 be assigned respectively  from the same number of independent cosets in $\hat{\Gamma}_{\mathbf{0}}-\hat{{\cal C}}$,
 associated to which the $2k$ ${\cal S}^{\varsigma_p}_{\kappa_p}$
 form an independent set of generators of $su(2^k)$,
 ${\cal S}^{\zeta_p}_{\mathbf{0}}\in{su(2^{n-k})}$.
 On account of the equality
 \begin{align}\label{eqcommnQ=kQ}
 [{\cal S}^{\zeta_p}_{\mathbf{0}}\otimes{\cal S}^{\varsigma_p}_{\kappa_p},\hspace{2pt}{\cal S}^{\zeta_q}_{\mathbf{0}}\otimes{\cal S}^{\varsigma_q}_{\kappa_q}]
 ={\cal S}^{\zeta_p+\zeta_q}_{\mathbf{0}}\otimes
 [{\cal S}^{\varsigma_p}_{\kappa_p},\hspace{2pt}{\cal S}^{\varsigma_q}_{\kappa_q}]
 \end{align}
 for all $1\leq p,q\leq 2k$,
 the generators ${\cal S}^{\varsigma_p}_{\kappa_p}$ span a Lie algebra isomorphic to $su(2^k)$.
 Moreover, among the $2k$ independent spinors ${\cal S}^{\zeta_p}_{\mathbf{0}}\otimes{\cal S}^{\varsigma_p}_{\kappa_p}$,
 it allows multiple options of a set of $k$ commuting members,
 for instance
 $\{ {\cal S}^{\zeta_l}_{\mathbf{0}}\otimes{\cal S}^{\varsigma_l}_{\mathbf{0}}:1\leq l\leq k \}$,
 to span a Cartan subalgebra of $su(2^n)$ with $\hat{{\cal C}}$~\cite{QAPSu0,QAPSu1}.
 \end{proof}
 \vspace{6pt}
 \vspace{6pt}
 \begin{lem}\label{intrinQAP}
 The QAP generated by the intrinsic bi-subalgebra $\hat{{\cal C}}$
 of the $k$-th maximum in the intrinsic Cartan subalgebra $\mathfrak{C}_{[\mathbf{0}]}\subset su(2^n)$
 is schematically illustrated as follows,
 \vspace{7pt}
  \[\begin{array}{lcr}
 &
 \hspace{-90pt}
 \begin{array}{c}
 \hat{{\cal C}}
 =W^1_{\tau=\mathbf{0},\mu=\mathbf{0}}\\
 \hspace{60pt}=\{ {\cal S}^{\zeta}_{\mathbf{0}}\otimes{\cal S}^{\mathbf{0}}_{\mathbf{0}}:\zeta\in{Z^{n-k}_2}  \}
 \end{array}
 &
 \\
 \\
 \hspace{-62pt}
 \begin{array}{c}
 W^{\epsilon}_{\tau,\hspace{1pt}\mu}
 =\{\hspace{2pt} {\cal S}^{\eta}_{\tau}\otimes{\cal S}^{\varsigma}_{\kappa}:\\
 \hspace{92pt}\eta\in{Z^{n-k}_2},\hspace{2pt} \mu=\varsigma\circ\kappa,\\
 \hspace{98pt}  \eta\cdot\tau+\varsigma\cdot\kappa=1+\epsilon \hspace{8pt}\}
 \end{array}
 &
 &
 \hspace{-63pt}
 \begin{array}{c}
 W^{1+\epsilon}_{\tau,\hspace{1pt}\mu}
 =\{ \hspace{2pt}{\cal S}^{\xi}_{\tau}\otimes{\cal S}^{\varsigma}_{\kappa}:\\
 \hspace{95pt}\xi\in{Z^{n-k}_2},\hspace{2pt} \mu=\varsigma\circ\kappa,\\
 \hspace{104pt}  \xi\cdot\tau+\varsigma\cdot\kappa=\epsilon\hspace{25pt}\},
 \end{array}
 \end{array}\]
 \vspace{5pt}
 ${\cal S}^{\zeta}_{\mathbf{0}}$, ${\cal S}^{\eta}_{\tau}$
 and ${\cal S}^{\xi}_{\tau}\in su(2^{n-k})$,
 ${\cal S}^{\varsigma}_{\kappa}\in su(2^k)$ with ${\cal S}^{\mathbf{0}}_{\mathbf{0}}$ as the identity,
 and $W^0_{\tau=\mathbf{0},\hspace{1pt}\mu=\mathbf{0}}=\{0\}$ being a null subspace,
 $\epsilon\in{Z_2}$.
 \end{lem}
 \vspace{2pt}
 \begin{proof}
 This lemma is Theorem~\ref{QAPoversu2^n} in the intrinsic coordinate.
 For example, consider two noncommuting spinors
 ${\cal S}^{\eta}_{\tau}\otimes{\cal S}^{\varsigma}_{\kappa}\in W^{\epsilon}_{\tau,\hspace{1pt}\mu}$
 and
 ${\cal S}^{\xi}_{\upsilon}\otimes{\cal S}^{\pi}_{\omega}\in W^{\sigma}_{\upsilon,\hspace{1pt}\nu}$.
 The bi-additive
 ${\cal S}^{\eta+\xi}_{\tau+\upsilon}\otimes{\cal S}^{\varsigma+\pi}_{\kappa+\omega}$
 belongs to $W^{\epsilon+\sigma}_{\tau+\upsilon,\hspace{1pt}\mu+\nu}$
 because
 $[{\cal S}^{\eta}_{\tau}\otimes{\cal S}^{\varsigma}_{\kappa},{\cal S}^{\xi}_{\upsilon}\otimes{\cal S}^{\pi}_{\omega}]
 =2(-1)^{\xi\cdot\tau+\pi\cdot\kappa}{\cal S}^{\eta+\xi}_{\tau+\upsilon}\otimes{\cal S}^{\varsigma+\pi}_{\kappa+\omega}$,
 $\mu=\varsigma\circ\kappa$, $\nu=\pi\circ\omega$
 and
 $\epsilon+\sigma=(1+\eta\cdot\tau+\varsigma\cdot\kappa)+(1+\xi\cdot\upsilon+\pi\cdot\omega)
 =1+(\eta+\xi)\cdot(\tau+\upsilon)+(\varsigma+\pi)\cdot(\kappa+\omega)$.
 \end{proof}
 \vspace{6pt}

 An explicit duality of partitions of spinors and codewords is shown.
\vspace{6pt}
 \begin{lem}\label{IntrCoordFundaLemmaCodeword}
 Given an ordered set of a number $n-k$ of independent spinors
 $\hat{\mathbf{S}}_{\hat{{\cal C}}}=\{ {\cal S}^{\zeta_r}_{\mathbf{0}}\otimes{\cal S}^{\mathbf{0}}_{\mathbf{0}}:
 \zeta_r=\sigma_{r,1}\sigma_{r,2}\cdots\sigma_{r,n-k}\in{Z^{n-k}_2}
 \text{ and }\sigma_{ru}=\delta_{ru},\hspace{2pt}r,u=1,2,\cdots,n-k \}$
 in $\hat{{\cal C}}$,
 the space of $n$-qubit states admits a decomposition
 ${\cal H}=\bigoplus_{\tau\in{Z^{n-k}_2}}\hat{{\cal H}}_{\tau}$ of
 disjoint eigen-invariant subspaces of the intrinsic bi-subalgebra $\hat{{\cal C}}$,
 where each $\hat{{\cal H}}_{\tau}$ of syndrome $\tau$
 is spanned by the orthonormal basis $\{\ket{\tau}\otimes\ket{i}:i\in{Z^k_2}\}$
 of dimension $2^k$.
 \end{lem}
 \vspace{2pt}
 \begin{proof}
  This lemma is Theorem~\ref{fundbasiscodewords} in the intrinsic coordinate.
  Since
  ${\cal S}^{\zeta_r}_{\mathbf{0}}\otimes{\cal S}^{\mathbf{0}}_{\mathbf{0}}\ket{\tau}\otimes\ket{i}
  =(-1)^{\zeta_r\cdot\tau}\ket{\tau}\otimes\ket{i}$ through Eq.~\ref{spinoronstate}
  for each $\tau\in{Z^{n-k}_2}$,
  ${\cal S}^{\zeta_r}_{\mathbf{0}}\otimes{\cal S}^{\mathbf{0}}_{\mathbf{0}}$
  being the $r$-th member in the ordered set $\hat{\mathbf{S}}_{\hat{{\cal C}}}$,
  the basis $\{\ket{\tau}\otimes\ket{i}:i\in{Z^k_2}\}$
  spans a $2^k$-dimensional eigen-invariant subspace
  $\hat{{\cal H}}_{\tau}$ of $\hat{{\cal C}}$ with syndrome
  $\tau=\epsilon_1\epsilon_2\cdots\epsilon_{n-k}$,
  $\epsilon_r=\zeta_r\cdot\tau\in{Z_2}$.
  Every $\ket{\tau}\otimes\ket{i}\in\hat{{\cal H}}_{\tau}$
  is a corrupted state of syndrome $\tau\neq\mathbf{0}$,
  and $\ket{\mathbf{0}}\otimes\ket{i}$
  is a basis codeword of the codeword subspace
  $\hat{{\cal H}}_{\mathbf{0}}$.
  Furthermore,
  the duality $\hat{\Gamma}_{\upsilon}(\hat{{\cal H}}_{\tau})\subset\hat{{\cal H}}_{\tau+\upsilon}$
  is unveiled in this coordinate straightforwardly
  ${\cal S}^{\eta}_{\upsilon}\otimes{\cal S}^{\pi}_{\omega}\ket{\tau}\otimes\ket{i}
  =(-1)^{\eta\cdot(\tau+\upsilon)+\pi\cdot(i+\omega)}\ket{\tau+\upsilon}\otimes\ket{i+\omega}\in\hat{{\cal H}}_{\tau+\upsilon}$
  for every ${\cal S}^{\eta}_{\upsilon}\otimes{\cal S}^{\pi}_{\omega}\in\hat{\Gamma}_{\upsilon}$
  and $\ket{\tau}\otimes\ket{i}\in\hat{{\cal H}}_{\tau}$,
  {\em cf.} Lemmas~\ref{BiSubIntrCoord} and~\ref{BlocksIntrinsic-n-k}.
  Apparently,
  $\hat{\Gamma}_{\mathbf{0}}(\hat{{\cal H}}_{\mathbf{0}})\subset\hat{{\cal H}}_{\mathbf{0}}$
  and $\hat{\Gamma}_{\mathbf{0}}(\hat{{\cal H}}_{\tau})\subset\hat{{\cal H}}_{\tau}$.

  As indicated in the proof of Lemma~\ref{BlocksIntrinsic-n-k},
  a number $2k$ of spinors chosen respectively from the same number of
  independent cosets in $\hat{\Gamma}_{\mathbf{0}}-\hat{{\cal C}}$
  generate a Lie algebra isomorphic to $su(2^k)$.
  Alternatively, the identity
  ${\cal S}^{\zeta}_{\mathbf{0}}\otimes{\cal S}^{\varsigma}_{\kappa}\ket{\mathbf{0}}\otimes\ket{i}
  =(-1)^{\varsigma\cdot(i+\kappa)}\ket{\mathbf{0}}\otimes\ket{i+\kappa}$
  via Eq.~\ref{spinoronstate} reaffirms the duty of
  ${\cal S}^{\zeta}_{\mathbf{0}}\otimes{\cal S}^{\varsigma}_{\kappa}\in\hat{\Gamma}_{\mathbf{0}}-\hat{{\cal C}}$
  to appear as an elementary operation of $k$ qubits.
 \end{proof}
 \vspace{6pt}
 The general form of orthogonality condition is spelled in the intrinsic coordinate.
 \vspace{6pt}
 \begin{lem}\label{OrthoCondIntrisicCoord}
 In the partition $[n,k,\hspace{1pt}\hat{{\cal C}}]$,
 the orthogonality is satisfied
  \begin{align}\label{orthocondintrinsic}
  \bra{\mathbf{0}}\otimes\bra{j}\hspace{2pt}
  {\cal S}^{\eta}_{\upsilon}\otimes{\cal S}^{\pi}_{\omega}
  \hspace{1pt}
  {\cal S}^{\zeta}_{\tau}\otimes{\cal S}^{\varsigma}_{\kappa}
  \hspace{2pt}\ket{\mathbf{0}}\otimes\ket{i}=(-1)^{\epsilon}\delta_{\upsilon\tau}\delta_{j+\omega,i+\kappa},
  \end{align}
 here
 $\epsilon=\zeta\cdot\tau+\eta\cdot\upsilon+\varsigma\cdot(i+\kappa)+\pi\cdot(j+\omega)$,
 $i,j\in{Z^k_2}$, $\tau,\upsilon,\zeta,\eta\in{Z^{n-k}_2}$ and $\kappa,\omega,\varsigma,\pi\in{Z^{k}_2}$.
 \end{lem}
 \vspace{2pt}
 \begin{proof}
 In the substitution of basis codewords and spinors $\ket{\psi_{\mathbf{0},i}}=\ket{\mathbf{0}}\otimes\ket{i}$,
 $\ket{\psi_{\mathbf{0},j}}=\ket{\mathbf{0}}\otimes\ket{j}$,
 $S_{\tau,\hspace{1pt}\mu}={\cal S}^{\zeta}_{\tau}\otimes{\cal S}^{\varsigma}_{\kappa}\in{\cal W}_{\tau,\hspace{1pt}\mu}$
 and
 $S_{\upsilon,\hspace{1pt}\nu}={\cal S}^{\eta}_{\upsilon}\otimes{\cal S}^{\pi}_{\omega}\in{\cal W}_{\upsilon,\hspace{1pt}\nu}$
 subscripted with coset indices $\mu=\varsigma\circ\kappa$ and $\nu=\pi\circ\omega$,
 Eq.~\ref{orthocondintrinsic} is the orthogonality condition
 of Corollary~\ref{OrthogonalityCond} in the intrinsic coordinate.
 If $\delta_{\upsilon\tau}=0$,
 the two entities $S_{\tau,\hspace{1pt}\mu}$ and $S_{\upsilon,\hspace{1pt}\nu}$
 are in distinct blocks.
 Whereas the two spinors are in a same coset of a block,
 if $\delta_{\upsilon\tau}=1$ and $\delta_{\mu\nu}=1$
 as $\varsigma=\pi$ and $\kappa=\omega$.
 The latter scenario reflects the code degeneracy implied by the concept of coset spinor,
 {\em cf.}  Lemma~\ref{CosetErrorCorrectable}.
 In regard to $\delta_{\upsilon\tau}=1$ and
 ${\cal S}^{\pi}_{\omega}\neq\pm{\cal S}^{\varsigma}_{\kappa}$,
 {\em i.e.}, $\nu\neq\mu$,
 the two spinors are in different cosets of a same block
 $\Gamma_{\tau\neq\mathbf{0}}$,
 namely errors uncorrectable, referring to Lemma~\ref{TwoCosetErrorUnCorrectable}.

 Despite in the intrinsic coordinate,
 Eq.~\ref{orthocondintrinsic} portraying the algebraic scenarios of the three cases respectively  is
 protected
 under a spinor-to-spinor mapping expounded in the next section
 and is thus generally true irrespective of coordinate choices.
 Importantly,
 it is easy to assert the equivalence of this orthogonality
 and the condition, necessary and sufficient, for the error correction in~\cite{KnillLaflamme}.
 \end{proof}
 \vspace{6pt}

 \section{Encoding}\label{secS-to-SMap}
 This section addresses the QAP transformation,
 basically a translation of Section~6 in~\cite{QAPSuTsai1}.
 \vspace{6pt}
 \begin{lem}\label{lemS-rot}
  An $s$-rotation ${\cal R}^{\zeta}_{\alpha}(\theta)=e^{i\theta(-i)^{\zeta\cdot\alpha}{\cal S}^{\zeta}_{\alpha}}\in{SU(2^n)}$
  of a spinor $(-i)^{\zeta\cdot\alpha}{\cal S}^{\zeta}_{\alpha}$
  has the expression
 \begin{align}\label{s-rots}
 e^{i\theta(-i)^{\zeta\cdot\alpha}{\cal S}^{\zeta}_{\alpha}}
 =cos\theta\hspace{1pt}{\cal S}^{\mathbf{0}}_{\mathbf{0}}+isin\theta\hspace{1pt}(-i)^{\zeta\cdot\alpha}{\cal S}^{\zeta}_{\alpha}
 \end{align}
  with the identity ${\cal S}^{\mathbf{0}}_{\mathbf{0}}\in{su(2^n)}$, $0\leq\theta <2\pi$.
 \end{lem}
 \vspace{2pt}
 \begin{proof}
  Refer to~\cite{QAPSu0}
  for the derivation.
 \end{proof}
 \vspace{6pt}
 The $s$-rotation
 $e^{i\theta(-i)^{\zeta\cdot\alpha}{\cal S}^{\zeta}_{\alpha}}$
 is an exponential transformation of $(-i)^{\zeta\cdot\alpha}{\cal S}^{\zeta}_{\alpha}$,
 a rotation acting on the space of $n$-qubit states,
 albeit not a rotation about the axis along this spinor.
 Remind that a spinor serves as an algebraic generator
 and also a group action.
 \vspace{6pt}
 \begin{lem}\label{lemS-rotStoSMap}
 An $s$-rotation ${\cal R}^{\zeta}_{\alpha}(\theta)\in{SU(2^n)}$
 is a spinor-to-spinor mapping as
 $\theta=\pm\frac{\pi}{2},\pm\frac{\pi}{4}$.
 \end{lem}
 \vspace{2pt}
 \begin{proof}
 By applying ${\cal R}^{\zeta}_{\alpha}(\theta)$ to a spinor $(-i)^{\eta\cdot\beta}{\cal S}^{\eta}_{\beta}$,
  it obtains
  \begin{align}\label{spintospinTrans}
  &{\cal R}^{\zeta\dag}_{\alpha}(\theta)\hspace{2pt}(-i)^{\eta\cdot\beta}{\cal S}^{\eta}_{\beta}\hspace{2pt}{\cal R}^{\zeta}_{\alpha}(\theta)\notag\\
  =&\hspace{10pt}(cos^2\theta+(-1)^{\eta\cdot\alpha+\zeta\cdot\beta}sin^2\theta)\hspace{2pt}(-i)^{\eta\cdot\beta}{\cal S}^{\eta}_{\beta}\notag\\
   &+\frac{i}{2}(-1)^{\zeta\cdot\beta}(-i)^{\zeta\cdot\alpha+\eta\cdot\beta}
    sin2\theta\hspace{1pt}(1-(-1)^{\eta\cdot\alpha+\zeta\cdot\beta}){\cal  S}^{\zeta+\eta}_{\alpha+\beta}.
 \end{align}
 For the detailed derivation, refer to~\cite{QAPSu0}.
 The operator remains invariant if $\eta\cdot\alpha+\zeta\cdot\beta=0,2$,
 that is,
 \begin{align}\label{spintospinTransComm}
  \hspace{-9pt}{\cal R}^{\zeta\dag}_{\alpha}(\theta)\hspace{2pt}(-i)^{\eta\cdot\beta}{\cal S}^{\eta}_{\beta}\hspace{2pt}{\cal R}^{\zeta}_{\alpha}(\theta)
 =(-i)^{\eta\cdot\beta}{\cal S}^{\eta}_{\beta}.
 \end{align}
 If $\eta\cdot\alpha+\zeta\cdot\beta=1,3$,
  this spinor is transformed into
 \begin{numcases}{({\cal R}^{\zeta}_{\alpha}(\theta)^{\dag}
 \hspace{2pt}(-i)^{\eta\cdot\beta}{\cal S}^{\eta}_{\beta}\hspace{2pt}
 {\cal R}^{\zeta}_{\alpha}(\theta)=}
  \hspace{46pt}-(-i)^{\eta\cdot\beta}\hspace{1pt}{\cal S}^{\eta}_{\beta}
  & as  \hspace{6pt}$\theta=\pm\frac{\pi}{2}$; \label{eqsRotpi/2}\\
  \pm \varrho\cdot(-i)^{(\zeta+\eta)\cdot(\alpha+\beta)}{\cal S}^{\zeta+\eta}_{\alpha+\beta}
  &  as $\hspace{6pt}\theta=\pm\frac{\pi}{4}$,\label{eqsRotpi/4}
\end{numcases}
 the coefficient
 $\varrho=i(-1)^{\zeta\cdot\beta}(-i)^{\zeta\cdot\alpha+\eta\cdot\beta}(i)^{(\zeta+\eta)\cdot(\alpha+\beta)}=\pm 1$.
 \end{proof}
 \vspace{6pt}
 The hermiticity
 is maintained by the arithmetic that
 the inner product $\zeta\cdot\alpha$ counts the number of
 the 1-qubit component ${\cal S}^1_1$ occurring in
 an $n$-qubit spinor ${\cal S}^{\zeta}_{\alpha}$,
 and the accumulated exponent $\zeta\cdot\alpha$
 of a phase $(\pm i)^{\zeta\cdot\alpha}$  is modulo $4$.
  \vspace{6pt}
 \begin{lem}\label{QpreserveBiadd}
  Every spinor-to-spinor mapping preserves the bi-addition
  respecting the relation
  \begin{align}\label{biaddsRelation}
  Q\hspace{1pt}(-i)^{(\xi+\mu)\cdot(\gamma+\nu)}{\cal S}^{\xi+\mu}_{\gamma+\nu}\hspace{1pt}Q^{\dag}
  =\sigma\cdot(-i)^{(\bar{\xi}+\bar{\mu})\cdot(\bar{\gamma}+\bar{\nu})}{\cal S}^{\bar{\xi}+\bar{\mu}}_{\bar{\gamma}+\bar{\nu}},
  \end{align}
  $\sigma=\pm 1$,
  for two spinors
  $(-i)^{\xi\cdot\gamma}{\cal S}^{\xi}_{\gamma}$ and $(-i)^{\mu\cdot\nu}{\cal S}^{\mu}_{\nu}\in{su(2^n)}$,
  and the pair
  $(-i)^{\bar{\xi}\cdot\bar{\gamma}}{\cal S}^{\bar{\xi}}_{\bar{\gamma}}=Q(-i)^{\xi\cdot\gamma}{\cal S}^{\xi}_{\gamma}Q^{\dag}$
  and
  $(-i)^{\bar{\mu}\cdot\bar{\nu}}{\cal S}^{\bar{\mu}}_{\bar{\nu}}=Q(-i)^{\mu\cdot\nu}{\cal S}^{\mu}_{\nu}Q^{\dag}$
  transformed by a spinor-to-spinor mapping $Q\in{SU(2^n)}$.
 \end{lem}
 \vspace{2pt}
 \begin{proof}
  This lemma is verified by equaling the two identities
 \begin{align}\label{biaddsBeta}
   &\hspace{0pt}Q(-i)^{\xi\cdot\gamma}{\cal S}^{\xi}_{\gamma}Q^{\dag}\cdot Q(-i)^{\mu\cdot\nu}{\cal S}^{\mu}_{\nu}Q^{\dag}\notag\\
  =& Q(-i)^{\xi\cdot\gamma}{\cal S}^{\xi}_{\gamma}\cdot(-i)^{\mu\cdot\nu}{\cal S}^{\mu}_{\nu}Q^{\dag}\notag\\
  =&\hspace{0pt}(-1)^{\mu\cdot\gamma}(-i)^{\xi\cdot\gamma+\mu\cdot\nu}(i)^{(\xi+\mu)\cdot(\gamma+\nu)}
               Q(-i)^{(\xi+\mu)\cdot(\gamma+\nu)}{\cal S}^{\xi+\mu}_{\gamma+\nu}Q^{\dag}
  \end{align}
  and
  \begin{align}\label{biaddsAlpha}
   \hspace{0pt}&Q(-i)^{\xi\cdot\gamma}{\cal S}^{\xi}_{\gamma}Q^{\dag}\cdot Q(-i)^{\mu\cdot\nu}{\cal S}^{\mu}_{\nu}Q^{\dag}\notag\\
  =\hspace{0pt}&(-i)^{\bar{\xi}\cdot\bar{\gamma}}{\cal S}^{\bar{\xi}}_{\bar{\gamma}}
                \cdot(-i)^{\bar{\mu}\cdot\bar{\nu}}{\cal S}^{\bar{\mu}}_{\bar{\nu}}\notag\\
  =\hspace{0pt}&(-1)^{\bar{\mu}\cdot\bar{\gamma}}(-i)^{\bar{\xi}\cdot\bar{\gamma}+\bar{\mu}\cdot\bar{\nu}}(i)^{(\bar{\xi}+\bar{\mu})\cdot(\bar{\gamma}+\bar{\nu})}
    (-i)^{(\bar{\xi}+\bar{\mu})\cdot(\bar{\gamma}+\bar{\nu})}{\cal S}^{\bar{\xi}+\bar{\mu}}_{\bar{\gamma}+\bar{\nu}},
  \end{align}
  $\sigma=(-1)^{\mu\cdot\gamma+\bar{\mu}\cdot\bar{\gamma}}
  (i)^{\xi\cdot\gamma+\mu\cdot\nu}(-i)^{(\xi+\mu)\cdot(\gamma+\nu)}
  (-i)^{\bar{\xi}\cdot\bar{\gamma}+\bar{\mu}\cdot\bar{\nu}}
  (i)^{(\bar{\xi}+\bar{\mu})\cdot(\bar{\gamma}+\bar{\nu})}=\pm 1$.
 \end{proof}
 \vspace{6pt}
  \vspace{6pt}
 \begin{thm}\label{QpreserveQAP}
 Every spinor-to-spinor mapping is QAP preserving.
 \end{thm}
 \vspace{2pt}
 \begin{proof}
 Attributed to the bi-addition preserving asserted in Lemma~\ref{QpreserveBiadd},
 a spinor-to-spinor mapping is coset preserving and upholds the structure of bi-subalgebra partition.
 Commuting with the bi-subalgebra ${\cal C}$
 of a partition $[n,k,\hspace{1pt}{\cal C}]$,
 the seed block $\Gamma_{\mathbf{0}}$ is maintained by this transformation,
 and thus so is every other block,
 that is, block preserving.
 Moreover, the unitarity of the transformation protects the commutation relation of every pair of spinors.
 Hence, the QAP structure is preserved under a spinor-to-spinor mapping.
 \end{proof}
 \vspace{6pt}

  The succeeding assertion details the procedure of
  constructing spinor-to-spinor mappings for the QAP transformation.
\vspace{6pt}
 \begin{thm}\label{s-to-smap2nindepspins}
  Given two ordered sets of $2n$ independent spinors
  $\mathbf{S}_1=\{ (-i)^{\xi_p\cdot\gamma_p}{\cal S}^{\xi_p}_{\gamma_p} \}$ and
  $\mathbf{S}_2=\{ (-i)^{\mu_q\cdot\nu_q}{\cal S}^{\mu_q}_{\nu_q} \}\subset{su(2^n)}$
  sharing the identical commutation relations
  $\xi_p\cdot\gamma_q+\xi_q\cdot\gamma_p=\mu_p\cdot\nu_q+\mu_q\cdot\nu_p$
  for the $p$-th spinor
  ${\cal S}^{\xi_p}_{\gamma_p}$
  and  the $q$-th ${\cal S}^{\xi_q}_{\gamma_q}$
  in $\mathbf{S}_1$
  as well as
  ${\cal S}^{\mu_p}_{\nu_p}$
  and
  ${\cal S}^{\mu_q}_{\nu_q}$
  in $\mathbf{S}_2$,
  there exists a spinor-to-spinor transformation $Q\in{SU(2^n)}$
  mapping the $p$-th member in $\mathbf{S}_1$
  to her counterpart in $\mathbf{S}_2$,
  i.e.,
  $(-i)^{\mu_p\cdot\nu_p}{\cal S}^{\mu_p}_{\nu_p}
  =Q^{\dag}(-i)^{\xi_p\cdot\gamma_p}{\cal S}^{\xi_p}_{\gamma_p}Q$,
  $p,q=1,2,\cdots,2n$.
 \end{thm}
 \vspace{2pt}
 \begin{proof}
  The key is to alter spinors in the ordered set  $\mathbf{S}_1$
  one by one but keep the preceding members invariant.
  In brief,
  after applications of the first $p-1$ evolutions $Q_r$ of the sequential mapping $Q=Q_1Q_2\cdots Q_{2n}$,
  $1\leq r<p\leq 2n$,
  by which
  the $r$-th member ${\cal S}^{\xi_r}_{\gamma_r}\in\mathbf{S}_1$
  is converted into ${\cal S}^{\mu_r}_{\nu_r}\in\mathbf{S}_2$
  and
  the $p$-th
  ${\cal S}^{\xi_p}_{\gamma_p}\in\mathbf{S}_1$ into $\pm{\cal S}^{\hspace{2pt}\iota_p}_{\varpi_p}$,
  the $p$-th operation $Q_p$ maps $\pm{\cal S}^{\hspace{2pt}\iota_p}_{\varpi_p}$
  to ${\cal S}^{\mu_p}_{\nu_p}\in\mathbf{S}_2$ and safeguards the preceding ${\cal S}^{\mu_r}_{\nu_r}$.
  Being unitary and formed in $s$-rotations ${\cal R}^{\xi}_{\gamma}(\theta)$
  as of Eqs.~\ref{eqsRotpi/2} and~\ref{eqsRotpi/4},
  every operation $Q_p$
  preserves the commutation relations
  among the $2n$ spinors respectively in
  $\mathbf{S}_1$ and $\mathbf{S}_2$.
  For convenience without confusion, ${\cal S}^{\xi}_{\gamma}$
  denotes a hermitian spinor by ignoring the phase $(-i)^{\xi\cdot\gamma}$.

  To build $Q_p$,
  three occasions are considered.
  The pair of spinors
  ${\cal S}^{\hspace{2pt}\iota_p}_{\varpi_p}$ and ${\cal S}^{\mu_p}_{\nu_p}$
  are identical apart from a sign $\pm 1$ on the 1st occasion,
  anticommuting secondly,
  and on the 3rd commuting but unequal.
  As $p=1$,
  the linear equations $\varsigma_p\cdot\nu_r+\mu_r\cdot\tau_p=0$
  of the parity constraint are of no effect owing to the absence of preceding ${\cal S}^{\mu_r}_{\nu_r}$.

  On the 1st occasion,
  there have a number $2^{2n-p}$ of solutions
  $Q_p={\cal R}^{\varsigma_p}_{\tau_p}(\frac{\pi}{2})$
  under the condition consisting of a number $p$ of linearly independent equations of parity constraint I
  $\varsigma_p\cdot\nu_p+\mu_p\cdot\tau_p=1$ and $\varsigma_p\cdot\nu_r+\mu_r\cdot\tau_p=0$
  based on Eq.~\ref{eqcommanticommspinors},
  such that ${\cal S}^{\mu_p}_{\nu_p}=Q^{\dag}_p(-{\cal S}^{\mu_p}_{\nu_p})Q_p$
  and ${\cal S}^{\mu_r}_{\nu_r}=Q^{\dag}_p{\cal S}^{\mu_r}_{\nu_r}Q_p$ according to Eq.~\ref{eqsRotpi/2},
  $1\leq r<p\leq 2n$.
  On the 2nd occasion,
  the operation $Q_p={\cal R}^{\iota_p+\mu_p}_{\varpi_p+\nu_p}(\pm\frac{\pi}{4})$
  turns $\pm{\cal S}^{\hspace{2pt}\iota_p}_{\varpi_p}$ into ${\cal S}^{\mu_p}_{\nu_p}$
  by Eq.~\ref{eqsRotpi/4} with an appropriate angle $\pm\frac{\pi}{4}$
  and fixes the preceding spinors ${\cal S}^{\mu_r}_{\nu_r}$
  given the conditions
  $\mu_r\cdot(\varpi_p+\nu_p)+(\iota_p+\mu_p)\cdot\nu_r=0$
  granted from the assumption of identical commutation relations
  $\mu_r\cdot\varpi_p+\iota_p\cdot\nu_r
  =\mu_r\cdot\nu_p+\mu_p\cdot\nu_r$.

  Now, come to $Q_{p}=Q_{p1}Q_{p2}$ on the 3rd occasion.
  There exist a number $2^{2n-p-1}$ of candidates
  $Q_{p1}={\cal R}^{\varsigma_p}_{\tau_p}(\frac{\pi}{4})$
  as in Eq.~\ref{eqsRotpi/4},
  each of which transforms $\pm{\cal S}^{\hspace{2pt}\iota_p}_{\varpi_p}$ to $\pm{\cal S}^{\iota_p+\varsigma_p}_{\varpi_p+\tau_p}$
  and leaves the preceding ${\cal S}^{\mu_r}_{\nu_r}$ unamended
  thanks to a number $p+1$ of linearly independent equations of parity constraint II
  \hspace{2pt}$\varsigma_p\cdot\varpi_p+\iota_p\cdot\tau_p=1$, $\varsigma_p\cdot\nu_p+\mu_p\cdot\tau_p=1$
  and $\varsigma_p\cdot\nu_r+\mu_r\cdot\tau_p=0$, $1\leq r<p\leq 2n$, {\em cf.} Eq.~\ref{eqcommanticommspinors}.
  Successively,
  the spinor
  $\pm{\cal S}^{\iota_p+\varsigma_p}_{\varpi_p+\tau_p}$
  is mapped to ${\cal S}^{\mu_p}_{\nu_p}$
  via
  $Q_{p2}={\cal R}^{\iota_p+\varsigma_p+\mu_p}_{\varpi_p+\tau_p+\nu_p}(\pm\frac{\pi}{4})$
  with an appropriate angle $\pm\frac{\pi}{4}$
  due to
  $\mu_p\cdot(\varpi_p+\tau_p+\nu_p)+(\iota_p+\varsigma_p+\mu_p)\cdot\nu_p=1$.
  Since
  $\mu_r\cdot(\varpi_p+\tau_p+\nu_p)+(\iota_p+\varsigma_p+\mu_p)\cdot\nu_r=0$,
  the preceding ${\cal S}^{\mu_r}_{\nu_r}$ remain intact.

  At the $(2n-1)$-th step, only the $3$rd occasion needs to be concerned, $p=2n-1\equiv  w$.
  Although there are in total $2n$ linear equations subject to the parity constraint II,
  one of them is derivable from the other independent $2n-1$.
  To assert this fact, it requires showing that
  ${\cal S}^{\iota_w}_{\varpi_w}$ belongs
  to the bi-subalgebra $\mathcal{B}$
  spanned by the $2n-1$ independent spinors
  ${\cal S}^{\mu_{r'}}_{\nu_{r'}}$, $1\leq r'< w$, and ${\cal S}^{\mu_w}_{\nu_w}$.
  Assume that ${\cal S}^{\iota_w}_{\varpi_w}\notin\mathcal{B}$.
  The identities
  $(\iota_w+\mu_w)\cdot\nu_{r'}+\mu_{r'}\cdot(\varpi_w+\nu_w)=0$
  are ascribed to the commutation relations
  $\iota_w\cdot\nu_{r'}+\mu_{r'}\cdot\varpi_w=\mu_w\cdot\nu_{r'}+\mu_{r'}\cdot\nu_w$.
  Next, in view of the commuting of
  ${\cal S}^{\iota_w}_{\varpi_w}$ and ${\cal S}^{\mu_w}_{\nu_w}$,
  there attain the two identities
  $(\iota_w+\mu_w)\cdot\varpi_w+\iota_w\cdot(\varpi_w+\nu_w)=0$
  and
  $(\iota_w+\mu_w)\cdot\nu_w+\mu_w\cdot(\varpi_w+\nu_w)=0$.
  Accordingly, these identities induce the implication that
  the bi-additive ${\cal S}^{\iota_w+\mu_w}_{\varpi_w+\nu_w}$
  commutes with the $2n$ independent spinors
  $\{  {\cal S}^{\mu_{r'}}_{\nu_{r'}},{\cal S}^{\mu_w}_{\nu_w},{\cal S}^{\iota_w}_{\varpi_w}:
  1\leq r'< w\}$,
  namely
  ${\cal S}^{\iota_w+\mu_w}_{\varpi_w+\nu_w}=\pm{\cal S}^{\mathbf{0}}_{\mathbf{0}}$,
  which contradicts the assumption ${\cal S}^{\iota_w}_{\varpi_w}\neq \pm{\cal S}^{\mu_w}_{\nu_w}$.
  This affirms ${\cal S}^{\iota_w}_{\varpi_w}\in\mathcal{B}$
  and the fact that the equation
  $\varsigma_w\cdot\varpi_w+\iota_w\cdot\tau_w=1$
  of constraint II can be derived from the other independent $2n-1$.
  As a result, the evolution $Q_{w1}$ of $Q_{w}=Q_{w1}Q_{w2}$ has two solutions.

  Finally, at step $2n$,
  the operation $Q_{2n}$ on the first two occasions respectively is produced
  through the same procedure as at the $p$-th step.
  On the 3rd occasion,
  the identities
  $(\iota_{2n}+\nu_{2n})\cdot\nu_{r''}+\mu_{r''}\cdot(\varpi_{2n}+\nu_{2n})=0$,
  $1\leq r''<2n$,
  are earned from the commutation relations
  $\iota_{2n}\cdot\nu_{r''}+\mu_{r''}\cdot\varpi_{2n}
  =\mu_{2n}\cdot\nu_{r''}+\mu_{r''}\cdot\nu_{2n}$.
  In addition, the assumption of the commuting
  ${\cal S}^{\iota_{2n}}_{\varpi_{2n}}$ and ${\cal  S}^{\mu_{2n}}_{\nu_{2n}}$
  yields the identity
  $(\iota_{2n}+\nu_{2n})\cdot\nu_{2n}+\mu_{2n}\cdot(\varpi_{2n}+\nu_{2n})=0$.
  Likewise, these $2n$ identities
  lead to the fact
  ${\cal S}^{\iota_{2n}+\mu_{2n}}_{\varpi_{2n}+\nu_{2n}}=\pm{\cal S}^{\mathbf{0}}_{\mathbf{0}}$.
  That is, the 3rd occasion reduces to the 1st.
  \end{proof}
 \vspace{6pt}
  \vspace{6pt}
 \begin{cor}\label{2nSindinnkC}
  Given an ordered set of $n-k$ independent spinors
  $\mathbf{S}'_{{\cal C}}=\{ {\cal S}^{\xi_g}_{\gamma_g}:g=1,2,\cdots,n-k \}$
  from a bi-subalgebra
  ${\cal C}\subset{su(2^n)}$,
  there exist a number $k$ of spinors
  $\{ {\cal S}^{\xi_m}_{\gamma_m}:m=n-k+1,\cdots,n \}$
  in $su(2^n)-{\cal C}$
  to span a Cartan subalgebra $\mathfrak{C}$ with $\mathbf{S}'_{{\cal C}}$,
  and a number $n$ of ${\cal S}^{\mu_t}_{\nu_t}\in{su(2^n)-\mathfrak{C}}$
  are independent under the bi-addition iff
  the associated strings
  $\omega_t=\epsilon_{t1}\epsilon_{t2}\cdots\epsilon_{tn}$
  are independent under the bitwise addition,
  $\epsilon_{ts}=\mu_t\cdot\gamma_s+\xi_s\cdot\nu_t\in{Z_2}$
  being the parities arising from the commutability of ${\cal S}^{\mu_t}_{\nu_t}$
  with the former $n$ members ${\cal S}^{\xi_s}_{\gamma_s}$,
  $s,t=1,2,\cdots, n$.
 \end{cor}
 \vspace{2pt}
 \begin{proof}
  By solving a number $n-k$ of independent linear equations
  of parities
  $\xi_m\cdot\gamma_{g}+\xi_{g}\cdot\gamma_m=0$ for all
  ${\cal S}^{\xi_g}_{\gamma_g}\in\mathbf{S}'_{{\cal C}}$,
  $1\leq g\leq n-k$ and $n-k+1\leq m\leq n$,
  there exist multiple solutions of $k$ independent spinors
  ${\cal S}^{\xi_m}_{\gamma_m}$
  to span a Cartan subalgebra $\mathfrak{C}$
  with $\mathbf{S}'_{{\cal C}}$,
  referring to~\cite{QAPSu0,QAPSu1} or the proof of Lemma~\ref{BlocksIntrinsic-n-k}.
  It is evident that every bi-additive ${\cal S}^{\mu_t+\mu_h}_{\nu_t+\nu_h}$
  of ${\cal S}^{\mu_t}_{\nu_t}$ and ${\cal S}^{\mu_h}_{\nu_h}$
  corresponds to the bitwise addition
  $\omega_t+\omega_h$ of two strings $\omega_t$ and $\omega_h$,
  $1\leq t,h\leq n$.
  Thus, the implication is validated that the set of latter $n$ spinors
  is independent iff so is the set of their associated strings.

  A preferred selection of an ordered set of $2n$ independent spinors contains all ${\cal S}^{\xi_s}_{\gamma_s}$
  and the 2nd half ${\cal S}^{\mu_t}_{\nu_t}$
  composed of $k$ members from the same number of independent cosets in
  $\Gamma_{\mathbf{0}}-{\cal C}$ and $n-k$ independent members
  respectively from distinct blocks $\Gamma_{\tau\neq\mathbf{0}}$.
  A such ordered set in the partition $[n,k,\hspace{1pt}\hat{{\cal C}}]$
  of the intrinsic coordinate is suggested,
  {\em cf.} Lemma~\ref{BlocksIntrinsic-n-k},
  \begin{align}\label{int2nIndpSpin}
  \hat{\mathbf{S}}
  =\{ & \hat{S}_{\hspace{1pt}l}={\cal S}^{\zeta_l}_{\mathbf{0}}\otimes{\cal S}^{\mathbf{0}}_{\mathbf{0}},
      \hspace{2pt}
      \hat{S}_{n-k+u}={\cal S}^{\mathbf{0}}_{\mathbf{0}}\otimes{\cal S}^{\varsigma_u}_{\mathbf{0}},
      \hspace{2pt}
      \hat{S}_{n+u}={\cal S}^{\mathbf{0}}_{\mathbf{0}}\otimes{\cal S}^{\mathbf{0}}_{\kappa_u}, \notag\\
      &\hat{S}_{n+k+l}={\cal S}^{\mathbf{0}}_{\tau_l}\otimes{\cal S}^{\mathbf{0}}_{\mathbf{0}}:
      l=1,2,\cdots,n-k,\text{ and }u=1,2,\cdots ,k \},
  \end{align}
  here $\zeta_{\hspace{1pt}l}=\varrho_{l,1}\varrho_{l,2}\cdots\varrho_{l,n-k}\in{Z^{n-k}_2}$
  obeying $\varrho_{\hspace{1pt}l,l'}=\delta_{\hspace{1pt}ll'}$
  for $1\leq l'\leq n-k$,
  $\varsigma_u=\iota_{u,1}\iota_{u,2}\cdots\iota_{u,k}\in{Z^k_2}$ fulfilling $\iota_{u,u'}=\delta_{uu'}$ for $1\leq u'\leq k$,
  $\kappa_{u}=\varsigma_{u-k+1}$
  and $\tau_l=\zeta_{n-k-l+1}$.
  The first $n-k$ independent members $\hat{S}_{\hspace{1pt}l}$ are picked from $\hat{{\cal C}}$,
  the $2k$ $\hat{S}_{n-k+u}$ and $\hat{S}_{n+u}$ from independent cosets in
  $\hat{\Gamma}_{\mathbf{0}}-\hat{{\cal C}}$,
  and the rest $n-k$ $\hat{S}_{n+k+l}$ from independent blocks $\Gamma_{\tau_l\neq\mathbf{0}}$.
  In this set, the first $n$ spinors span the intrinsic Cartan subalgebra of $su(2^n)$.
  As to the last $n$ members,
  each $\hat{S}_{n+u}$ is associated with the string
  $\omega_u=\epsilon_1\epsilon_2\cdots\epsilon_n\in{Z^n_2}$
  carrying $\epsilon_{n-u+1}=1$ and $\epsilon_h=0$ if $h\neq n-u+1$,
  and every $\hat{S}_{n+k+l}$ of the string
  $\omega_{k+l}=\varrho_1\varrho_2\cdots\varrho_n$
  possessing $\varrho_{n-k-l+1}=1$ and $\varrho_h=0$ if $h\neq n-k-l+1$.
 \end{proof}
 \vspace{6pt}
 Maneuvering a QAP transformation begins with the assignment
 $\mathbf{S}_2=\hat{\mathbf{S}}$ of the suggested set
 of Eq.~\ref{int2nIndpSpin} in the intrinsic coordinate.
 Let the other ordered set $\mathbf{S}_1$ of $2n$ independent spinors
 that enjoys the identical commutation relations as those of $\mathbf{S}_2$
 and respects the independence condition of
 Corollary~\ref{2nSindinnkC}
 be drawn from the partition $[n,k,\hspace{1pt}{\cal C}]$
 by solving linear equations of parities of
 Theorem~\ref{s-to-smap2nindepspins},
 referring to the proof of
 Corollary~\ref{QcmapsEncoding} for
 its existence.
 According to Theorem~\ref{s-to-smap2nindepspins} again,
 there provide a number $2n$ of sequential spinor-to-spinor
 operations
 $\{Q^{\dag}_{p}:p=1,2,\cdots, 2n\}$
 recursively changing each member of $\mathbf{S}_1$
 to her opposite in $\mathbf{S}_2$,
 and thus mapping $[n,k,\hspace{1pt}{\cal C}]$
 to $[n,k,\hspace{1pt}\hat{{\cal C}}]$.

 Nonetheless,
 it costs less to transmute
 $[n,k,\hspace{1pt}\hat{{\cal C}}]$
 to
 $[n,k,\hspace{1pt}{\cal C}]$
 simply by turning $\hat{{\cal C}}$ into ${\cal C}$,
 namely applying only the last $n-k$ evolutions $Q_{n-k}$, $\cdots$, $Q_{1}$ of the sequential
 operations $\{Q_{\hat{p}}:\hat{p}=2n,2n-1,\cdots,1\}$, which is regarded as an {\em encoding}~\cite{SuTsaiOptFTQC}.
 \vspace{6pt}
 \begin{cor}\label{QcmapsEncoding}
 Transforming the partition $[n,k,\hspace{1pt}\hat{{\cal C}}]$
 into $[n,k,\hspace{1pt}{\cal C}]$
 is achievable by an encoding $Q_{en}\in{SU(2^n)}$
 that converts the intrinsic bi-subalgebra $\hat{{\cal C}}$
 to ${\cal C}=Q_{en}\hspace{1pt}\hat{{\cal C}}\hspace{1pt}Q^{\dag}_{en}$.
 \end{cor}
 \vspace{2pt}
 \vspace{2pt}
 \begin{proof}
 Reminisce in Lemma~\ref{IntrCoordFundaLemmaCodeword} that
 the ordered set
 $\hat{\mathbf{S}}_{\hat{{\cal C}}}
 =\{ {\cal S}^{\zeta_r}_{\mathbf{0}}\otimes{\cal S}^{\mathbf{0}}_{\mathbf{0}}:
 r=1,2,\cdots,n-k \}$
 is designated from $\hat{{\cal C}}$.
 Endowed with the truth that a partition is determined by its generating bi-subalgebra,
 an encoding transforms
 $[n,k,\hspace{1pt}\hat{{\cal C}}]$ into $[n,k,\hspace{1pt}{\cal C}]$
 by converting $\hat{{\cal C}}$ to ${\cal C}$,
 specifically with the procedure of Theorem~\ref{s-to-smap2nindepspins} mapping $\hat{\mathbf{S}}_{\hat{{\cal C}}}$
 of $\hat{{\cal C}}$ spinor by spinor
 to an independent ordered set $\mathbf{S}'_{{\cal C}}$ of
 ${\cal C}$ as in Corollary~\ref{2nSindinnkC}.
 The immense freedom of selecting $\mathbf{S}'_{{\cal C}}$
 from ${\cal C}$
 gives a combinatorially large number of all versions of encodings
 ${\cal N}=\prod^{n-k-1}_{l=0}(2^{n-k}-2^{l})$.
 Meanwhile, there have at least the number ${\cal N}$ of candidates of an independent ordered set $\mathbf{S}_1$
 sharing the identical commutation relations
 with those of $\hat{\mathbf{S}}$,
 {\em cf.} Corollary~\ref{2nSindinnkC},
 for each of which is easily acquired by mapping $\hat{\mathbf{S}}$ to the partition $[n,k,\hspace{1pt}{\cal C}]$
 via an encoding.
 \end{proof}
 \vspace{6pt}
 After exercising the inverse
 of the encoding $Q^{\dag}_{en}=Q^{\dag}_{n-k}\cdots Q^{\dag}_1$,
 each coset ${\cal W}_{\tau,\hspace{1pt}\mu}$ of ${\cal C}$
 is altered to a coset $\hat{{\cal W}}_{\tau',\hspace{1pt}\mu'}$ of $\hat{{\cal C}}$ with the syndrome $\tau'\in{Z^{n-k}_2}$.
 Whereas, every syndrome retains $\tau'=\tau$ under $Q^{\dag}_{en}$
 if the ordered set $\mathbf{S}'_{{\cal C}}\subset\mathbf{S}_1$
 is identical to
 $\mathbf{S}_{{\cal C}}$ comprising
 the detection operators appointed from ${\cal C}$, {\em cf.} Definition~\ref{defsyndrome}.
 This alignment $\mathbf{S}'_{{\cal C}}=\mathbf{S}_{{\cal C}}$ is assumed in the following exposition.
 The coset $\hat{{\cal W}}_{\tau,\hspace{1pt}\mu'}$ of coset index $\mu'$
 further switches to $\hat{{\cal W}}_{\tau,\hspace{1pt}\mu}$ of index $\mu$
 through the next $2k$ evolutions
 $Q^{\dag}_{n+k}\cdots Q^{\dag}_{n-k+1}$.
 And, finally, each of the last $n-k$ $\{Q^{\dag}_{p'}:p'=n+k+1,n+k+2, \cdots, 2n\}$ permutes spinors
 in a same coset~\cite{SuTsaiOptFTQC}.

 \section{Fault Tolerance}\label{secFTQC}
 The formulation of fault tolerant encodes is guided by two criteria.
\vspace{6pt}
\begin{prop}\label{propFTaction}
 For the partition $[n,k,\hspace{1pt}{\cal C}]$ generated by a bi-subalgebra ${\cal C}\subset su(2^n)$,
 an action $U\in SU(2^n)$ is fault tolerant by fulfilling two criteria,
 the eigen-invariance, i.e.,
 $SU\ket{\psi}=U\ket{\psi}$ for each spinor $S\in{\cal C}$
 and every codeword $\ket{\psi}$,
 and the error correction against an error set ${\cal E}$,
 i.e.,
 $US_{\beta}\ket{\psi}
 =\sum_{\hspace{1pt}\alpha\in\hspace{1pt}{Z^{n-k}_2}-\{\mathbf{0}\}}x_{\alpha\beta}S_{\alpha,\hspace{1pt}\nu}\hspace{1pt}U\ket{\psi}$
 for $S_{\beta}\in{\cal E}$
 with $S_{\alpha,\hspace{1pt}\nu}$ in a coset
 ${\cal W}_{\alpha,\hspace{1pt}\nu}$ uniquely from each block $\Gamma_{\alpha}$,
 $\nu\in{Z^{2k}_2}$ and $x_{\alpha\beta}\in\mathbb{C}$.
\end{prop}
\vspace{6pt}
 The criterion of eigen-invariance obliges
 $U\ket{\psi}$
 to stay as a codeword of ${\cal C}$,
 to which the partition $[n,k,\hspace{1pt}{\cal C}]$
 is applicable.
 The 2nd criterion is to pledge
 every error of ${\cal E}$ correctable
 by $[n,k,\hspace{1pt}{\cal C}]$,
 that is, referring to Theorem~\ref{ErrorSetcorrectable},
 a corruption
 $US_{\beta}\ket{\psi}$
 affected by an error $S_{\beta}\in{\cal E}$
 admits a linear expansion in correctable states
 $S_{\alpha,\hspace{1pt}\nu}\hspace{1pt}U\ket{\psi}$,
 here $S_{\alpha,\hspace{1pt}\nu}$ being an arbitrary spinor
 from a single coset ${\cal W}_{\alpha,\hspace{1pt}\nu}$
 in $\Gamma_{\alpha}$.
 The derivation is conducted in the intrinsic coordinate.
\vspace{6pt}
 \begin{lem}\label{EigenInvIntrinsic}
 In the partition $[n,k,\hspace{1pt}\hat{{\cal C}}]$ generated by the intrinsic bi-subalgebra $\hat{{\cal C}}$,
 an eigen-invariant action $\hat{U}\in{SU(2^n)}$ takes the block-diagonal form
\begin{align}\label{formEigenInvAction}
 \hat{U}=\Lambda\oplus\Omega
\end{align}
 with $\Lambda=\ket{\mathbf{0}}\bra{\mathbf{0}}\otimes M_{\mathbf{0},\mathbf{0}}$,
 $M_{\mathbf{0},\mathbf{0}}\in{SU(2^k)}$,
 $\Omega=\sum_{\alpha,\hspace{1pt}\beta\hspace{1pt}\in\hspace{1pt}{Z^{n-k}_2}-\{\mathbf{0}\}}
 \ket{\alpha}\bra{\beta}\otimes M_{\alpha,\beta}$,
 $M_{\alpha,\beta}\in\mathbb{C}^{2^k\times 2^k}$,
 and $\ket{\mathbf{0}}$, $\ket{\alpha}$ and $\ket{\beta}$ being respectively a basis state
 of $n-k$ qubits.
 \end{lem}
 \vspace{2pt}
 \begin{proof}
  An eigen-invariant action $\hat{U}$ is required to satisfy the
  condition
\begin{align}\label{CondofEigen}
 {\cal S}^{\zeta}_{\mathbf{0}}\otimes{\cal S}^{\mathbf{0}}_{\mathbf{0}}\hspace{2pt} \hat{U}\ket{\mathbf{0}}\otimes \ket{i}=\hat{U}\ket{\mathbf{0}}\otimes \ket{i}
\end{align}
 for each spinor ${\cal S}^{\zeta}_{\mathbf{0}}\otimes{\cal S}^{\mathbf{0}}_{\mathbf{0}}\in\hat{{\cal C}}$
 and every basis codeword  $\ket{\mathbf{0}}\otimes \ket{i}$, $\zeta\in{Z^{n-k}_2}$ and $i\in{Z^k_2}$.
 Let a $2^n\times 2^n$ matrix be cast as
 \begin{align}\label{HyperMatAction}
 \hat{U}=\ket{\mathbf{0}}\bra{\mathbf{0}}\otimes M_{\mathbf{0},\mathbf{0}}
 +\sum_{\alpha\hspace{1pt}\neq\mathbf{0}}\ket{\alpha}\bra{\mathbf{0}}\otimes M_{\alpha,\mathbf{0}}
 +\sum_{\beta\neq\mathbf{0}}\ket{\mathbf{0}}\bra{\beta}\otimes M_{\mathbf{0},\beta}
 +\sum_{\alpha,\hspace{1pt}\beta\neq\mathbf{0}}\ket{\alpha}\bra{\beta}\otimes
 M_{\alpha,\beta},
 \end{align}
 $M_{\tau,\upsilon}\in\mathbb{C}^{2^k\times 2^k}$
 and $\ket{\tau}\in\mathbb{C}^{2^{n-k}}$ being an $(n-k)$-qubit basis state,
 $\tau,\upsilon\in{Z^{n-k}_2}$.
 Equating LHS
\begin{align}\label{LHSEigen}
 {\cal S}^{\zeta}_{\mathbf{0}}\otimes{\cal S}^{\mathbf{0}}_{\mathbf{0}}\hspace{2pt}\hat{U}\ket{\mathbf{0}}\otimes\ket{i}
 =\ket{\mathbf{0}}\otimes M_{\mathbf{0},\mathbf{0}}\ket{i}
 +\sum_{\alpha\neq\mathbf{0}}(-1)^{\zeta\cdot\alpha}\ket{\alpha}\otimes M_{\alpha,\mathbf{0}}\ket{i}
\end{align}
 and RHS
\begin{align}\label{RHSEigen}
 \hat{U}\ket{\mathbf{0}}\otimes\ket{i}
 =\ket{\mathbf{0}}\otimes M_{\mathbf{0},\mathbf{0}}\ket{i}
 +\sum_{\alpha\neq\mathbf{0}}\ket{\alpha}\otimes
 M_{\alpha,\mathbf{0}}\ket{i}
\end{align}
 of Eq.~\ref{CondofEigen} results in
\begin{align}\label{LHS=RHSEigen}
 \sum_{\alpha\neq\mathbf{0}}\{(-1)^{\zeta\cdot\alpha}-1\}\ket{\alpha}\otimes M_{\alpha,\mathbf{0}}\ket{i}=0.
\end{align}
 With respect to each basis state $\ket{\alpha}$,
 the equality
\begin{align}\label{LHS=RHSEigenEachTerm}
 \{(-1)^{\zeta\cdot\alpha}-1\}\ket{\alpha}\otimes M_{\alpha,\mathbf{0}}\ket{i}=0
\end{align}
 is attained for all $\zeta\in{Z^{n-k}_2}$ and $i\in{Z^k_2}$.
 Since, for every $\alpha\neq\mathbf{0}$,
 there exists a string $\zeta\in{Z^{n-k}_2}$
 enjoying $\zeta\cdot\alpha=1$,
 in general $-2\ket{\alpha}\otimes M_{\alpha,\mathbf{0}}\ket{i}\neq 0$
 unless $M_{\alpha,\mathbf{0}}=\mathbf{O}$, the zero matrix.

 Hence, the matrix $\hat{U}$ is put into
\begin{align}\label{EqCalEigenUNotUni}
 \hat{U}=\ket{\mathbf{0}}\bra{\mathbf{0}}\otimes M_{\mathbf{0},\mathbf{0}}
 +\sum_{\beta\neq \mathbf{0}}\ket{\mathbf{0}}\bra{\beta}\otimes M_{\mathbf{0},\beta}
 +\sum_{\alpha,\beta\neq {\bf 0}}\ket{\alpha}\bra{\beta}\otimes
 M_{\alpha,\beta}.
\end{align}
 In virtue of
 $\hat{U}^{\dag}\hat{U}=\hat{U}\hat{U}^{\dag}=I_{2^n}$,
 the matrix $\hat{U}$ is settled as
 \begin{align}\label{eqfinalUhat}
  \hat{U}=\ket{\mathbf{0}}\bra{\mathbf{0}}\otimes M_{\mathbf{0},\mathbf{0}}
 +\sum_{\alpha,\beta\neq {\bf 0}}\ket{\alpha}\bra{\beta}\otimes M_{\alpha,\beta}.
 \end{align}
 By denoting
 $\Omega=\sum_{\alpha,\hspace{1pt}\beta\neq\mathbf{0}}
 \ket{\alpha}\bra{\beta}\otimes M_{\alpha,\beta}$,
 the unitarity
 \begin{align}\label{OmegDagOmegUnitary}
 \Omega^{\dag}\Omega=\sum_{\beta,\beta'\neq\mathbf{0}}\ket{\beta'}\bra{\beta}\otimes
  \sum_{\alpha\neq\mathbf{0}} M^{\dag}_{\alpha,\beta'}M_{\alpha,\beta}=I_{2^n-2^k}
 \end{align}
  and
 \begin{align}\label{OmegOmegDagUnitary}
 \Omega\Omega^{\dag}=\sum_{\alpha,\alpha'\neq\mathbf{0}}\ket{\alpha}\bra{\alpha'}\otimes
  \sum_{\beta\neq\mathbf{0}} M_{\alpha,\beta}M^{\dag}_{\alpha',\beta}=I_{2^n-2^k}
 \end{align}
 is drawn.
 With further denoting the operation
 $\Lambda=\ket{\mathbf{0}}\bra{\mathbf{0}}\otimes M_{\mathbf{0},\mathbf{0}}$,
 an eigen-invariant action of the block-diagonal form
 $\hat{U}=\Lambda\oplus\Omega$ is acquired.
 It is apparent that the product $\hat{U}_2\hat{U}_1$ of two eigen-invariant actions
 $\hat{U}_1=\Lambda_1\oplus\Omega_1$ and $\hat{U}_2=\Lambda_2\oplus\Omega_2$
 remains eigen-invariant owing to
 $\hat{U}_2\hat{U}_1=(\Lambda_2\oplus\Omega_2)\cdot(\Lambda_1\oplus\Omega_1)
 =\Lambda_2\Lambda_1\oplus\Omega_2\Omega_1$.
 \end{proof}
 \vspace{6pt}
 The eigen-invariance leaves great freedom of choosing  $\Omega$
 that grants the redundancy demanded in the error correction as shown in Lemma~\ref{FTactionIntFprm}.
 The next two assertions confirms the function authenticity
 of the operator $\hat{U}=\Lambda\oplus\Omega$.
 \vspace{6pt}
 \begin{lem}\label{Authentics}
 Given a mapping
 \begin{align}\label{IntrM00=Aj}
 M_{\mathbf{0},\mathbf{0}}\ket{i}
 =\sum_{j\in{Z^k_2}}a_{ji}\ket{j}
 \end{align}
 of an operation $M_{\mathbf{0},\mathbf{0}}\in{SU(2^k)}$
 acting on a basis state $\ket{i}$ of $k$ qubits,
 the same form
 \begin{align}\label{UlinearEigenAj}
 U\ket{\bar{i}}=\sum_{j\in{Z^k_2}}a_{ji}\ket{\bar{j}}
 \end{align}
 holds for the operation $U=Q\hat{U}Q^{\dag}$ acting on the basis state
 $\ket{\bar{i}}=Q\ket{\mathbf{0}}\otimes\ket{i}$
 of $n$ qubits via a unitary transformation $Q\in{SU(2^n)}$,
 here $\ket{\bar{j}}=Q\ket{\mathbf{0}}\otimes\ket{j}$ and $\hat{U}=\Lambda\oplus\Omega\in{SU(2^n)}$
 being block diagonal,
 $\Lambda= \ket{\mathbf{0}}\bra{\mathbf{0}}\otimes M_{\mathbf{0},\mathbf{0}}$,
 $\Omega=\sum_{\alpha,\beta\neq\mathbf{0}}\ket{\alpha}\bra{\beta}\otimes M_{\alpha,\beta}$,
 $M_{\alpha,\beta}\in\mathbb{C}^{2^k\times 2^k}$,
 and $\ket{\mathbf{0}}$, $\ket{\alpha}$ and $\ket{\beta}$ being a basis state
 of $n-k$ qubits.
 \end{lem}
 \vspace{2pt}
 \begin{proof}
 The validity of this lemma is attributed to the fact
 $\hat{U}\ket{\mathbf{0}}\otimes\ket{i}=\sum_{j}a_{ji}\ket{\mathbf{0}}\otimes\ket{j}$
 in the intrinsic coordinate, and
 performing $Q$ is simply a coordinate transformation.
\end{proof}
\vspace{6pt}
 \vspace{6pt}
\begin{lem}\label{faithFunSpinors}
 Given an equality $M_2M_1\ket{i}=M_3\ket{i}$ for three unitary operations
 $M_s\in{SU(2^k)}$ and a basis state $\ket{i}$
 of $k$ qubits, $s=1,2,3$,
 the same form
 $U_2U_1\ket{\bar{i}}=U_3\ket{\bar{i}}$
 is preserved for the operations $U_s=Q\hat{U}_sQ^{\dag}$ and the basis state
 $\ket{\bar{i}}=Q\ket{\mathbf{0}}\otimes\ket{i}$
 of $n$ qubits via a unitary transformation $Q\in{SU(2^n)}$,
 here
 $\hat{U}_s=\Lambda_s\oplus\Omega_s\in{SU(2^n)}$
 being block diagonal,
 $\Lambda_s=\ket{\mathbf{0}}\bra{\mathbf{0}}\otimes M_s$,
 $\Omega_s=\sum_{\alpha,\beta\neq\mathbf{0}}\ket{\alpha}\bra{\beta}\otimes M^{(s)}_{\alpha,\beta}$,
 $M^{(s)}_{\alpha,\beta}\in\mathbb{C}^{2^k\times 2^k}$,
 and $\ket{\mathbf{0}}$, $\ket{\alpha}$ and $\ket{\beta}$ being a basis state
 of $n-k$ qubits.
\end{lem}
\vspace{2pt}
\begin{proof}
 Similarly,
 this lemma will be asserted in the intrinsic coordinate.
 Notice that
 in general $\hat{U}_2\hat{U}_1\neq \hat{U}_3$ unless
 $M^{(3)}_{\alpha,\beta}=\sum_{\gamma\in{Z^{n-k}_2}}M^{(2)}_{\alpha,\gamma}M^{(1)}_{\gamma,\beta}$.
 However, the relation is true as actions applied to
 codewords,
 namely
\begin{align}\label{eqRelPreservekOps}
  &\hat{U}_2\hat{U}_1\ket{\mathbf{0}}\otimes\ket{i}\notag\\
 =&\{\ket{\mathbf{0}}\bra{\mathbf{0}}\otimes M_2+\sum_{\alpha,\beta\neq\mathbf{0}}\ket{\alpha}\bra{\beta}\otimes M^{(2)}_{\alpha,\beta}\}
   \{\ket{\mathbf{0}}\bra{\mathbf{0}}\otimes M_1+\sum_{\alpha',\beta'\neq\mathbf{0}}\ket{\alpha'}\bra{\beta'}\otimes M^{(1)}_{\alpha',\beta'}\}
   \hspace{2pt}\ket{\mathbf{0}}\otimes\ket{i}\notag\\
 =&\{\ket{\mathbf{0}}\bra{\mathbf{0}}\otimes M_3
   +\sum_{\alpha,\beta\neq \mathbf{0}}\ket{\alpha}\bra{\beta}\otimes M^{(3)}_{\alpha,\beta}\}\hspace{2pt}\ket{\mathbf{0}}\otimes\ket{i}\notag\\
 =&\hat{U}_3\ket{\mathbf{0}}\otimes\ket{i}.
\end{align}
\end{proof}
 \vspace{6pt}

 An eigen-invariant action correcting errors is derived.
\vspace{6pt}
 \begin{lem}\label{FTactionIntFprm}
 In the partition $[n,k,\hspace{1pt}\hat{{\cal C}}]$,
 an eigen-invariant action
 $\hat{U}=\Lambda\oplus\Omega$
 associated to a $k$-qubit action $M_{\mathbf{0},\mathbf{0}}\in{SU(2^k)}$,
 $\Lambda=\ket{\mathbf{0}}\bra{\mathbf{0}}\otimes M_{\mathbf{0},\mathbf{0}}$
 and $\Omega=\sum_{\alpha,\beta\in{Z^{n-k}_2-\{\mathbf{0}\}}}\ket{\alpha}\bra{\beta}\otimes M_{\alpha,\beta}$,
 is error correctable if each
 $M_{\alpha,\beta}\in{\mathbb{C}^{2^k\times 2^k}}$
 is of the form
\begin{align}\label{FTactionMab}
 M_{\alpha,\beta}=i^{\xi_{\alpha\beta}\cdot\alpha}(-i)^{\eta_{\beta}\cdot\beta}x_{\alpha\beta}\textsf{S}_{\alpha}M_{\mathbf{0},\mathbf{0}}\mathbb{S}_{\beta},
\end{align}
 where
 the $k$-qubit spinors
 $\mathbb{S}_{\beta}=(-i)^{\varsigma_{\beta}\cdot\kappa_{\beta}}{\cal S}^{\varsigma_{\beta}}_{\kappa_{\beta}}$
 and
 $\textsf{S}_{\alpha}=(-i)^{\pi_{\alpha}\cdot\omega_{\alpha}}{\cal S}^{\pi_{\alpha}}_{\omega_{\alpha}}$
 are respectively the input coset
 $\hat{{\cal W}}_{\beta,\hspace{1pt}\mu(\beta)}$ of block $\hat{\Gamma}_{\beta}$
 and the output coset
 $\hat{{\cal W}}_{\alpha,\hspace{1pt}\nu(\alpha)}$ of block $\hat{\Gamma}_{\alpha}$,
 $\mu(\beta)=\varsigma_{\beta}\circ\kappa_{\beta}$
 and
 $\nu(\alpha)=\pi_{\alpha}\circ\omega_{\alpha}$,
 and the unitarity
 $\widetilde{{\cal T}}\widetilde{{\cal T}}^{\dag}=I_{2^n-2^k}$
 and
 $\widetilde{{\cal T}}^{\dag}\widetilde{{\cal T}}=I_{2^n-2^k}$
 holds for the transfer amplitude
 $\widetilde{{\cal T}}=\sum_{\alpha,\beta\neq\mathbf{0}}
 i^{\xi_{\alpha\beta}\cdot\alpha}(-i)^{\eta_{\beta}\cdot\beta}x_{\alpha\beta}
 \ket{\alpha}\bra{\beta}\otimes I_{2^k}$,
 $x_{\alpha\beta}\in\mathbb{C}$ and $\xi_{\alpha\beta},\eta_{\beta}\in{Z^{n-k}_2}$.
 \end{lem}
 \vspace{2pt}
 \begin{proof}
 Assume given a partition $[n,k,\hspace{1pt}{\cal C}]$ against a correctable error set
 ${\cal E}=\{ E^{(\tau')}:\tau'\in{J}\subseteq{Z^{n-k}_2} \}$,
 {\em cf.} Theorem~\ref{ErrorSetcorrectable},
 composed of spinors from distinct blocks $\Gamma_{\tau'}$,
 $J$ being a subset of $Z^{n-k}_2$.
 By Corollary~\ref{QcmapsEncoding}, there exists the adjoint of an encoding $Q^{\dag}_{en}\in{SU(2^n)}$
 mapping each error $E^{(\tau)}\in\Gamma_{\tau}$
 into a spinor
 $S^{(\tau)}=Q^{\dag}_{en}\hspace{1pt}E^{(\tau)}\hspace{1pt}Q_{en}
 =(-i)^{\eta_{\tau}\cdot\tau}{\cal S}^{\eta_{\tau}}_{\tau}
  \otimes (-i)^{\varsigma_{\tau}\cdot\kappa_{\tau}}{\cal S}^{\varsigma_{\tau}}_{\kappa_{\tau}}\in
  \hat{{\cal W}}_{\tau,\hspace{1pt}\vartheta}$,
 as of Lemma~\ref{BlocksIntrinsic-n-k},
 in $[n,k,\hspace{1pt}\hat{{\cal C}}]$ of the intrinsic coordinate.
 The coset $\hat{{\cal W}}_{\tau,\hspace{1pt}\vartheta}\subset\hat{\Gamma}_{\tau}$
 containing  $S^{(\tau)}$
 is reckoned as a coset of errors
 and subscripted with the coset index
 $\vartheta=\varsigma_{\tau}\circ\kappa_{\tau}$.
 Let the part of $k$ qubits of $S^{(\tau)}$ be denoted as
 $\mathbb{S}_{\tau}=(-i)^{\varsigma_{\tau}\cdot\kappa_{\tau}}{\cal S}^{\varsigma_{\tau}}_{\kappa_{\tau}}$
 and identified with the coset
 $\hat{{\cal W}}_{\tau,\hspace{1pt}\vartheta}$,
 again referring to Lemma~\ref{BlocksIntrinsic-n-k}.
 Following the prescription of Theorem~\ref{ErrorSetcorrectable},
 the $k$-qubit part $\mathbb{S}_{\tau}$
 is further recognized as an {\em input coset}
 $\hat{{\cal W}}_{\beta,\hspace{1pt}\mu(\beta)}$
 solely designated in block $\hat{\Gamma}_{\beta}$,
 $\mu(\beta)=\varsigma_{\beta}\circ\kappa_{\beta}$,
 {\em i.e.}, $\hat{{\cal W}}_{\tau,\hspace{1pt}\vartheta}=\hat{{\cal W}}_{\beta,\hspace{1pt}\mu(\beta)}$
 as $\tau=\beta$.
 Towards the purpose of error correction, it requires a set of input cosets
 $\mathcal{P}_{in}
 =\{ \hat{{\cal W}}_{\beta,\hspace{1pt}\mu(\beta)}:\beta\in{Z^{n-k}_2} \}
 =\{\mathbb{S}_{\beta}:\beta\in{Z^{n-k}_2}\}$
 consisting of a unique coset chosen from every block
 as well as being a superset of cosets of errors.

 Suppose a corruption
 affected by an error
\begin{align}\label{EactiontoCorruptSt}
 \hat{U}S^{(\beta)}\ket{\mathbf{0}}\otimes\ket{i}
 =\sum_{\alpha\neq \mathbf{0}}
 \ket{\alpha}\otimes
 \{i^{\eta_{\beta}\cdot\beta}M_{\alpha,\beta}\mathbb{S}_{\beta}\}\ket{i}.
\end{align}
 If $\hat{U}$ is error correctable,
 according to Proposition~\ref{propFTaction},
 this corruption allows the expression
\begin{align}\label{FormofCorrect}
 \hat{U}S^{(\beta)}\ket{\mathbf{0}}\otimes\ket{i}
 =\sum_{\alpha\neq \mathbf{0}}x_{\alpha\beta}\hspace{1pt}S_{\alpha\hookleftarrow\beta}\hspace{1pt}\hat{U}\ket{\mathbf{0}}\otimes\ket{i}
 =\sum_{\alpha\neq \mathbf{0}}
  \ket{\alpha}\otimes
  \{i^{\xi_{\alpha\beta}\cdot\alpha}x_{\alpha\beta}\textsf{S}_{\alpha}M_{\mathbf{0},\mathbf{0}}\}\ket{i}.
\end{align}
 Likewise,
 for the error correction,
 each spinor
 $S_{\alpha\hookleftarrow\beta}
 =(-i)^{\xi_{\alpha\beta}\cdot\alpha}{\cal S}^{\xi_{\alpha\beta}}_{\alpha}
 \otimes(-i)^{\pi_{\alpha}\cdot\omega_{\alpha}}{\cal S}^{\pi_{\alpha}}_{\omega_{\alpha}}$
 individually belongs to an {\em output} coset
 $\hat{{\cal W}}_{\alpha,\hspace{1pt}\nu(\alpha)}$
 uniquely designated in $\hat{\Gamma}_{\alpha}$,
 $\nu(\alpha)=\pi_{\alpha}\circ\omega_{\alpha}$.
 Abbreviated as
 $\textsf{S}_{\alpha}=(-i)^{\pi_{\alpha}\cdot\omega_{\alpha}}{\cal S}^{\pi_{\alpha}}_{\omega_{\alpha}}$
 the $k$-qubit part of $S_{\alpha\hookleftarrow\beta}$
 is identified with the output coset $\hat{{\cal W}}_{\alpha,\hspace{1pt}\nu(\alpha)}$ solely in $\hat{\Gamma}_{\alpha}$,
 and there constitutes a set of output cosets
 $\mathcal{P}_{out}=\{ \hat{{\cal W}}_{\alpha,\hspace{1pt}\nu(\alpha)}:\alpha\in{Z^{n-k}_2} \}
 =\{\textsf{S}_{\alpha}:\alpha\in{Z^{n-k}_2}\}$.
 Note that $S_{\alpha\hookleftarrow\beta}$
 is no necessary correlated to the input spinor
 $S^{(\beta)}$,
 {\em e.g.}, the phase
 $\xi_{\alpha\beta}=\xi_{\alpha}$
 depending only on the block $\hat{\Gamma}_{\alpha}$.

  By equating Eqs.~\ref{EactiontoCorruptSt}
  and~\ref{FormofCorrect},
  the operator responsible for error corrections is deduced
\begin{align}\label{termMalbefinal}
 M_{\alpha,\beta}
  =i^{\xi_{\alpha\beta}\cdot\alpha}(-i)^{\eta_{\beta}\cdot\beta}x_{\alpha\beta}\textsf{S}_{\alpha}M_{\mathbf{0},\mathbf{0}}\mathbb{S}_{\beta}.
\end{align}
  The requirement
  $\Omega\Omega^{\dag}=I_{2^n-2^k}$
  and
  $\Omega^{\dag}\Omega=I_{2^n-2^k}$ of $\Omega$ in Lemma~\ref{EigenInvIntrinsic}
 leads to the unitarity  of amplitudes
\begin{align}\label{UnitaCondAmp}
 &\sum_{\beta\neq\mathbf{0}}
 (i^{\xi_{\alpha\beta}\cdot\alpha}(-i)^{\eta_{\beta}\cdot\beta}x_{\alpha\beta})
 \hspace{2pt}
 (i^{\xi_{\alpha'\beta}\cdot\alpha'}(-i)^{\eta_{\beta}\cdot\beta}x_{\alpha'\beta})^*
 =\delta_{\alpha\alpha'}
  \hspace{6pt}\text{ and}\notag\\
 &\sum_{\alpha\neq\mathbf{0}}
 (i^{\xi_{\alpha\beta'}\cdot\alpha}(-i)^{\eta_{\beta'}\cdot\beta'}x_{\alpha\beta'})^*
 (i^{\xi_{\alpha\beta}\cdot\alpha}(-i)^{\eta_{\beta}\cdot\beta}x_{\alpha\beta})
 =\delta_{\beta'\beta}
\end{align}
  invoking the properties $\textsf{S}^\dag_{\alpha}=\textsf{S}_{\alpha}$,
 $\mathbb{S}^\dag_{\beta}=\mathbb{S}_{\beta}$,
 $\textsf{S}^2_{\alpha}=I_{2^k}$,
 $\mathbb{S}^2_{\beta}=I_{2^k}$,
 and
 $M_{\mathbf{0},\mathbf{0}}M^\dag_{\mathbf{0},\mathbf{0}}=M^\dag_{\mathbf{0},\mathbf{0}}M_{\mathbf{0},\mathbf{0}}
 =I_{2^k}$.
 The unitarity of Eq.~\ref{UnitaCondAmp}
 is briefed as $\widetilde{{\cal T}}\widetilde{{\cal T}}^{\dag}=I_{2^n-2^k}$
 and $\widetilde{{\cal T}}^{\dag}\widetilde{{\cal T}}=I_{2^n-2^k}$
 by formulating the transfer amplitude
 $\widetilde{{\cal T}}=\sum_{\alpha,\beta\neq\mathbf{0}}
 i^{\xi_{\alpha\beta}\cdot\alpha}(-i)^{\eta_{\beta}\cdot\beta}x_{\alpha\beta}
 \ket{\alpha}\bra{\beta}\otimes I_{2^k}$.
 In the incidence of no correlation $\xi_{\alpha\beta}=\xi_{\alpha}$,
 Eq.~\ref{UnitaCondAmp}  reduces to
 \begin{align}\label{UnitaCondAmpSumInd}
 &\sum_{\beta\neq\mathbf{0}}x_{\alpha\beta} x^*_{\alpha'\beta}
 =\delta_{\alpha\alpha'}
  \hspace{2pt}\text{ and }\hspace{2pt}
 \sum_{\alpha\neq\mathbf{0}} x^*_{\alpha\beta'}x_{\alpha\beta}
 =\delta_{\beta'\beta}.
\end{align}
 To ensure the unitarity of the {\em correction} operator $\Omega$,
  the sets of input cosets
 $\mathcal{P}_{in}$
 and output cosets
 $\mathcal{P}_{out}$
 need be formed {\em in completeness}.
 That is, either of the complete sets is constructed by taking
 one, and only one, coset from every block.
 The introduction of
 $\mathcal{P}_{in}$
 and
 $\mathcal{P}_{out}$,
 as a consequence of identifying the $k$-qubit part of a spinor with a
 coset, {\em cf.} Lemma~\ref{BlocksIntrinsic-n-k},
 is explicitly an articulation of the concept of coset spinor.
 By default, the two complete sets are no necessarily produced in correlation.
 A major clue is to prepare ${\cal P}_{in}$ as a superset of cosets of errors.
 It is a plausible view that the cosets of ${\cal P}_{in}$
 is transformed into the cosets of ${\cal P}_{out}$ through a {\em rotation}
 woven from the transfer amplitude
 $\widetilde{{\cal T}}$.
 In building $\Omega$,
 there is vast freedom to decide the two coset sets ${\cal P}_{in}$ and ${\cal P}_{out}$,
 and write the transfer amplitude
 $\widetilde{{\cal T}}$.

 Once more an echo of the concept of coset spinor,
 an identical $M_{\alpha,\beta}$ is obtained if two inputs
 $S^{(\beta)}_{a}=(-i)^{\eta_a\cdot\beta}{\cal S}^{\eta_a}_{\beta}\otimes\mathbb{S}_{\beta}$
 and
 $S^{(\beta)}_{b}=(-i)^{\eta_b\cdot\beta}{\cal S}^{\eta_b}_{\beta}\otimes\mathbb{S}_{\beta}$
 are in a same abelian subspace of a coset $\hat{{\cal W}}_{\beta,\hspace{1pt}\mu(\beta)}$,
 or two versions apart from the global phase
 $i^{\eta_a\cdot\beta}(-i)^{\eta_b\cdot\beta}$
 if the two operators are in distinct abelian subspaces of $\hat{{\cal W}}_{\beta,\hspace{1pt}\mu(\beta)}$.
 A similar argument is applicable to two spinors in a same output coset as
 $\xi_{\alpha\beta}=\xi_{\alpha}$.

 Reexamine the corruption $\hat{U}\hat{E}_{in}\ket{\mathbf{0}}\otimes\ket{i}$
 infected with a noise
 $\hat{E}_{in}=\sum_{\beta\in{Z^{n-k}_2}}y_{\beta}S^{(\beta)}$,
 $y_{\beta}\in\mathbb{C}$ and $\sum_{\beta}|y_{\beta}|^2=1$,
 contributed by
 $S^{(\beta)}=(-i)^{\eta_{\beta}\cdot\beta}{\cal S}^{\eta_{\beta}}_{\beta}\otimes\mathbb{S}_{\beta}\in{\cal W}_{\beta,\hspace{1pt}\mu(\beta)}$
 from distinct blocks,
 $S^{(\mathbf{0})}=I_{2^n}$ and $\mathbb{S}_{\mathbf{0}}=I_{2^k}$.
 Let the corruption be detected by
 ${\cal S}^{\zeta}_{\mathbf{0}}\otimes{\cal S}^{\mathbf{0}}_{\mathbf{0}}\in\hat{{\cal C}}$,
 \begin{align}\label{examSdUSerrket}
  &{\cal S}^{\zeta}_{\mathbf{0}}\otimes{\cal S}^{\mathbf{0}}_{\mathbf{0}}\hat{U}\hat{E}_{in}\ket{\mathbf{0}}\otimes\ket{i}\notag\\
 =&{\cal S}^{\zeta}_{\mathbf{0}}\otimes{\cal S}^{\mathbf{0}}_{\mathbf{0}}
 \{ y_{\mathbf{0}}\ket{\mathbf{0}}\otimes M_{\mathbf{0},\mathbf{0}}\ket{i}
  +\sum_{\alpha\neq\mathbf{0}}(\sum_{\beta\neq\mathbf{0}}i^{\xi_{\alpha\beta}\cdot\alpha}\hspace{1pt}y_{\beta}  x_{\alpha\beta})
  \ket{\alpha}\otimes\textsf{S}_{\alpha}M_{\mathbf{0},\mathbf{0}}\ket{i}\}\notag\\
 =& y_{\mathbf{0}}\ket{\mathbf{0}}\otimes M_{\mathbf{0},\mathbf{0}}\ket{i}
  +\sum_{\alpha\neq\mathbf{0}}(-1)^{\zeta\cdot\alpha}(\sum_{\beta\neq\mathbf{0}}i^{\xi_{\alpha\beta}\cdot\alpha}\hspace{1pt}y_{\beta}x_{\alpha\beta})
  \ket{\alpha}\otimes\textsf{S}_{\alpha}M_{\mathbf{0},\mathbf{0}}\ket{i}.
 \end{align}
 The detection of Eq.~\ref{examSdUSerrket} is conducted
 in use of {\em detectors} supported by
 an ordered set of $n-k$ independent spinors in $\hat{{\cal C}}$,
 for instance, but no restricted to, the ordered set $\hat{\mathbf{S}}_{\hat{{\cal C}}}$ as of
 Lemma~\ref{IntrCoordFundaLemmaCodeword}.
 The non-error state
 $\hat{U}\ket{\mathbf{0}}\otimes\ket{i}=\ket{\mathbf{0}}\otimes M_{\mathbf{0},\mathbf{0}}\ket{i}$
 occurs with the probability $|y_{\mathbf{0}}|^2$.
 Subsequently a corrupted state
 $\ket{\alpha}\otimes\textsf{S}_{\alpha}M_{\mathbf{0},\mathbf{0}}\ket{i}$
 of syndrome $\alpha$ arises with the probability
 $|\sum_{\beta\neq\mathbf{0}}i^{\xi_{\alpha\beta}\cdot\alpha}\hspace{1pt}y_{\beta}x_{\alpha\beta}|^2$.
 The non-error state
 $(-i)^{\xi\cdot\alpha}\hspace{1pt}\ket{\mathbf{0}}\otimes M_{\mathbf{0},\mathbf{0}}\ket{i}$
 is recovered
 by applying a correction operator
 $\hat{S}_{\alpha,\hspace{1pt}\nu(\alpha)}
 =(-i)^{\xi\cdot\alpha}{\cal S}^{\xi}_{\alpha}\otimes\textsf{S}_{\alpha}$
 arbitrarily taken from the single output coset
 $\hat{{\cal W}}_{\alpha,\hspace{1pt}\nu(\alpha)}$
 in block $\hat{\Gamma}_{\alpha}$,
 $\nu(\alpha)=\pi_{\alpha}\circ\omega_{\alpha}\in{Z^{2k}_2}$
 and the part of $k$ qubits
 $\textsf{S}_{\alpha}=(-i)^{\pi_{\alpha}\cdot\omega_{\alpha}}{\cal S}^{\pi_{\alpha}}_{\omega_{\alpha}}$.

 Accompanied with a set of correction operators
 $\{S_{\alpha,\hspace{1pt}\nu(\alpha)}\in{\cal W}_{\alpha,\hspace{1pt}\nu(\alpha)}:\alpha\in{Z^{n-k}_2}\}$
 appointed from cosets
 ${\cal W}_{\alpha,\hspace{1pt}\nu(\alpha)}=Q_{en}\hspace{1pt}\hat{{\cal W}}_{\alpha,\hspace{1pt}\nu(\alpha)}\hspace{1pt}Q^{\dag}_{en}$,
 a fault tolerant action $U=Q_{en}\hspace{1pt}\hat{U}\hspace{1pt}Q^{\dag}_{en}$
 is introduced to the partition
 $[n,k,\hspace{1pt}{\cal C}]$
 via the encoding $Q_{en}\in{SU(2^n)}$, {\em cf.} Corollary~\ref{QcmapsEncoding}.
 \end{proof}
\vspace{6pt}
 The closure of fault tolerant actions is again a reflection of the concept of coset spinor.
\vspace{6pt}
 \begin{lem}\label{prod2FTactions}
 The product $U_2U_1$ of two fault tolerant encodes is fault tolerant
 if, as in a same block,
 the input coset of $U_2$
 is identical to the output coset of $U_1$.
 \end{lem}
 \vspace{2pt}
 \begin{proof}
   It suffices to assert this lemma in the intrinsic coordinate.
   Following Lemma~\ref{FTactionIntFprm},
   assume given two fault tolerant actions
   $\hat{U}_h=\Lambda_h\oplus\Omega_h$ associated with $M^{(h)}_{\mathbf{0},\mathbf{0}}\in{SU(2^k)}$,
   $h=1,2$,
   where
   $\Lambda_h=\ket{\mathbf{0}}\bra{\mathbf{0}}\otimes M^{(h)}_{\mathbf{0},\mathbf{0}}$,
   $\Omega_h=\sum_{\alpha,\beta\neq\mathbf{0}}\ket{\alpha}\bra{\beta}\otimes M^{(h)}_{\alpha,\beta}$,
   $M^{(h)}_{\alpha,\beta}=i^{\xi_{\alpha\beta}\cdot\alpha}(-i)^{\eta_{\beta}\cdot\beta}
   x^{(h)}_{\alpha\beta}\textsf{S}^{(h)}_{\alpha}M^{(h)}_{\mathbf{0},\mathbf{0}}\mathbb{S}^{(h)}_{\beta}\in\mathbb{C}^{2^k\times 2^k}$,
   and
   $\mathbb{S}^{(h)}_{\beta}\in\mathcal{P}^{(h)}_{in}$
   and $\textsf{S}^{(h)}_{\alpha}\in\mathcal{P}^{(h)}_{out}$ in $\Omega_h$
   are respectively the input coset of block $\hat{\Gamma}_{\beta}$
   and the output coset of block $\hat{\Gamma}_{\alpha}$.

   For an error
   $E_{\tau}=(-i)^{\eta_{\tau}\cdot\tau}{\cal S}^{\eta_{\tau}}_{\tau}\otimes\mathbb{S}_{\tau}$
   adhering to $\mathbb{S}_{\tau}=\mathbb{S}^{(1)}_{\beta}$ as
   $\tau=\beta\in{Z^{n-k}_2}$,
   the fault tolerance of the action $\hat{U}_2\hat{U}_1$
   applied to a codeword
  $\ket{\mathbf{0}}\otimes\ket{i}$
   is confirmed as follows, $i\in{Z^k_2}$,
  \begin{align}\label{FTclosure}
   &\hat{U}_2\hat{U}_1 E_{\tau}\ket{\mathbf{0}}\otimes\ket{i}\notag\\
  =&\hat{U}_2\hat{U}_1\{(-i)^{\eta_{\tau}\cdot\tau}{\cal S}^{\eta_{\tau}}_{\tau}\otimes\mathbb{S}_{\tau} \}\ket{\mathbf{0}}\otimes\ket{i}\notag\\
  =&
    \sum_{\alpha\neq\mathbf{0}}
    \ket{\alpha}\otimes
    \{ (\sum_{\beta\neq\mathbf{0}}i^{\xi_{\alpha\beta}\cdot\alpha}
    (-i)^{\eta_{\beta}\cdot\beta} i^{\xi_{\beta\tau}\cdot\beta}
    x^{(2)}_{\alpha\beta}x^{(1)}_{\beta\tau})
    \textsf{S}^{(2)}_{\alpha}M^{(2)}_{\mathbf{0},\mathbf{0}} M^{(1)}_{\mathbf{0},\mathbf{0}}\ket{i}\}\notag\\
   &\hspace{4pt}(\mathbb{S}^{(2)}_{\beta}=\textsf{S}^{(1)}_{\alpha'}
    \hspace{6pt}\text{ if in a same block } \Gamma_{\beta=\alpha'})\notag\\
  =&
    \sum_{\alpha\neq\mathbf{0}}\ket{\alpha}\otimes
    \{ i^{\xi_{\alpha\beta}\cdot\alpha}
    x_{\alpha\tau}\textsf{S}_{\alpha}M_{\mathbf{0},\mathbf{0}}\ket{i}\},
  \end{align}
  in the imposition of the criteria
  $M_{\mathbf{0},\mathbf{0}}=M^{(2)}_{\mathbf{0},\mathbf{0}}M^{(1)}_{\mathbf{0},\mathbf{0}}$,
  $\textsf{S}_{\alpha}=\textsf{S}^{(2)}_{\alpha}$,
  $x_{\alpha\tau}=\sum_{\beta\neq\mathbf{0}}
  x^{(2)}_{\alpha\beta}x^{(1)}_{\beta\tau}$
  and
  $(-i)^{\eta_{\beta}\cdot\beta}
  i^{\xi_{\beta\tau}\cdot\beta}=1$.
  The unitarity
  $\sum_{\tau\neq\mathbf{0}}x_{\alpha'\tau}x^{*}_{\alpha\tau}=\delta_{\alpha'\alpha}$
  and
  $\sum_{\alpha'\neq\mathbf{0}}x^{*}_{\alpha'\tau'}x_{\alpha'\tau}=\delta_{\tau'\tau}$
   is accordingly assured.
 \end{proof}
\vspace{6pt}

 The fault tolerance encode of an action is no unique.
\vspace{6pt}
 \begin{thm}\label{FTacionsforEach[nkt]}
 For the partition $[n,k,\hspace{1pt}{\cal C}]$ generated by a
 bi-subalgebra ${\cal C}\subset su(2^n)$ against an error set
 ${\cal E}$,
 there exist multiple choices of fault tolerant encodes for every action of $k$ qubits.
 \end{thm}
\vspace{2pt}
\begin{proof}
 Given $[n,k,\hspace{1pt}{\cal C}]$,
 the adjoint of an encoding $Q^{\dag}_{en}\in{SU(2^n)}$,
 by Corollary~\ref{QcmapsEncoding},
 transforms each error $E^{(\tau)}\in\Gamma_{\tau}$ of ${\cal E}$
 to a spinor
 $S^{(\tau)}=Q^\dag_{en}\hspace{1pt} E^{(\tau)}\hspace{1pt}Q_{en}=(-i)^{\eta_{\tau}\cdot\tau}{\cal S}^{\eta_{\tau}}_{\tau}\otimes\tilde{\mathbb{S}}_{\tau}$
 of a coset of errors $\hat{{\cal W}}_{\tau,\hspace{1pt}\vartheta}\subset\hat{\Gamma}_{\tau}$
 in $[n,k,\hspace{1pt}\hat{{\cal C}}]$
 of the intrinsic coordinate,
 $\vartheta=\varsigma_{\tau}\circ\omega_{\tau}$
 and the part of $k$ qubits
 $\tilde{\mathbb{S}}_{\tau}=(-i)^{\varsigma_{\tau}\cdot\omega_{\tau}}{\cal S}^{\varsigma_{\tau}}_{\omega_{\tau}}$,
 {\em cf.} Theorem~\ref{ErrorSetcorrectable} and Lemma~\ref{BlocksIntrinsic-n-k}.

 Let a fault tolerant action
 $\hat{U}=\Lambda\oplus\Omega$
 of the intrinsic coordinate
 as in Lemma~\ref{FTactionIntFprm}
 be prepared
 for a $k$-qubit action $M_{\mathbf{0},\mathbf{0}}\in{SU(2^k)}$,
 here
 $\Lambda=\ket{\mathbf{0}}\bra{\mathbf{0}}\otimes M_{\mathbf{0},\mathbf{0}}$,
 $\Omega=\sum_{\alpha,\beta\neq\mathbf{0}}\ket{\alpha}\bra{\beta}\otimes
 (i^{\xi_{\alpha\beta}\cdot\alpha}(-i)^{\eta_{\beta}\cdot\beta}x_{\alpha\beta}\textsf{S}_{\alpha}M_{\mathbf{0},\mathbf{0}}\mathbb{S}_{\beta})$,
 and $\mathbb{S}_{\beta}$ being the input coset $\hat{{\cal W}}_{\beta,\hspace{1pt}\mu(\beta)}\in\mathcal{P}_{in}$
 of block $\hat{\Gamma}_{\beta}$
 and $\textsf{S}_{\alpha}$ the output coset $\hat{{\cal W}}_{\alpha,\hspace{1pt}\nu(\alpha)}\in\mathcal{P}_{out}$
 of block $\hat{\Gamma}_{\alpha}$.
 As a must, the next is to identify the input coset
 with the coset of errors
 $\hat{{\cal W}}_{\beta,\hspace{1pt}\mu(\beta)}=\hat{{\cal W}}_{\tau,\hspace{1pt}\vartheta}$
 in every same block $\hat{\Gamma}_{\beta=\tau}$,
 {\em i.e.},
 $\mathbb{S}_{\beta}=\tilde{\mathbb{S}}_{\tau}$ if $\beta=\tau$, referring to Theorem~\ref{ErrorSetcorrectable}.
 Then,
 by mapping $\hat{U}$ into $U=Q_{en}\hspace{1pt}\hat{U}\hspace{1pt}Q^{\dag}_{en}$,
 an action of fault tolerance $U$ of $M_{\mathbf{0},\mathbf{0}}$ encoded in $[n,k,\hspace{1pt}{\cal C}]$ is delivered,
 accompanied with a set of correction operators
 $\{ S_{\alpha,\hspace{1pt}\nu(\alpha)}\in{\cal W}_{\alpha,\hspace{1pt}\nu(\alpha)}:\alpha\in{Z^{n-k}_2} \}$
 individually from each output coset
 ${\cal W}_{\alpha,\hspace{1pt}\nu(\alpha)}=Q_{en}\hspace{1pt}\hat{{\cal W}}_{\alpha,\hspace{1pt}\nu(\alpha)}\hspace{1pt}Q^{\dag}_{en}$.

 Multiple choices of fault tolerant encodes
 for a $k$-qubit action in $[n,k,\hspace{1pt}{\cal C}]$
 lie in great freedom of constructing the encoding
 $Q_{en}$ and the correction operatior $\Omega$.
 A huge number of versions of $Q_{en}$
 spring from rich options of selecting ordered sets of $n-k$ independent
 spinors from ${\cal C}$, referring to Corollary~\ref{QcmapsEncoding}.
 As explicated in Theorem~7 in~\cite{QAPSuTsai1},
 all bi-subalgebras sized the same as ${\cal C}$ are acquirable via exhaustive spinor-to-spinor mappings,
 amounting to a combinatorially gigantic number of partitions given $n$ and $k$.
 Beware the intractable complexity to further determine admissible partitions
 correcting an error set.
 The diversified design of $\Omega$
 stems from numerous designations of cosets for the two complete sets
 ${\cal P}_{in}$ and ${\cal P}_{out}$
 and from the extensive weaving of transfer amplitude
 $\widetilde{{\cal T}}$, {\em cf.} Lemma~\ref{FTactionIntFprm}.
 Cost deviations~\cite{QAPSuTsai2},
 differing in $Q_{en}$ and $\Omega$,
 escalate when the computation scaling up~\cite{SuTsaiOptFTQC}.
\end{proof}
\vspace{6pt}
 No confined to stabilizer codes,
 the methodology of creating fault tolerant encodes is also applicable
 to nonadditive codes through the {\em removing process}~\cite{SuTsaiOptFTQC}.
 In respect of a partition
 $[n,k,\hspace{1pt}{\cal C};t]$ correcting $t$-errors,
 {\em cf.} Corollary~\ref{t-errorCorrectable},
 the gap $n-k$ can be arbitrarily squeezed
 with increasing $n$ and $k$ given mildly rising $t$,
 namely
 the rate $R=\frac{k}{n}\rightarrow 1$
 for very large $n$ and $k$ given $t$~\cite{GottesmanStab}.
 In corporation with versatile selections of universal sets of gates
 in accord with constraints and advantages of implementations in practice,
 it is algorithmically achievable to attain optimized compositions of actions~\cite{QAPSu0,QAPSuTsai1,QAPSuTsai2}.
 Synthesizing the two elements paves the way to an Initiative of
 optimizing scalable fault tolerance quantum computation~\cite{NARLNCHC}.
 The assertion of conclusion may be the beginning of new adventures.
 \vspace{6pt}
 \begin{cor}\label{nkcodekqubitFT}
 Every action in every code admits fault tolerance.
 \end{cor}
 \vspace{6pt}

\nonumsection{References}

\end{document}